\documentclass[11pt,a4paper]{article}
\pdfoutput=1
\usepackage{jheppub}

\usepackage{dcolumn}
\usepackage{latexsym}

\newcommand{\be}{\begin{equation}}
\newcommand{\ee}{\end{equation}}
\newcommand{\bea}{\begin{eqnarray}}
\newcommand{\eea}{\end{eqnarray}}

\newcommand{\refc}[1]{(\ref{#1})}
\newcommand{\csw}{c_\text{SW}}
\newcommand{\msbar}{\overline{\text{MS}}}

\newcommand{\ve}{\mathbf}

\newcommand{\Tc}{T_C}
\newcommand{\Nt}{N_t}

\newcommand{\U}{\text{U}}
\newcommand{\SU}{\text{SU}}
\newcommand{\Ua}{\text{U}_{\rm A}(1)}

\renewcommand{\Re}{\text{Re}}
\newcommand{\Tr}{\text{Tr}}

\newcommand{\bare}{^{\rm bare}}
\newcommand{\ren}{^{\rm ren}}

\newcommand{\cond}{{\langle\bar{\psi}\psi\rangle}}

\newcommand{\scan}[1]{{\bf #1}}

\title{\boldmath On the strength of the $\Ua$ anomaly at the chiral
phase transition in $N_f=2$ QCD}

\author{Bastian B. Brandt$^{a,b}$, Anthony Francis$^c$, Harvey B. Meyer$^d$, Owe
Philipsen$^a$, Daniel Robaina$^d$, Hartmut Wittig$^d$}
\affiliation{
$^a$Institut f\"ur Theoretische Physik, Goethe-Universit\"at, D-60438 Frankfurt
am Main \\
$^b$Institut f\"ur theoretische Physik, Universit\"at Regensburg, D-93040
Regensburg \\
$^c$Department of Physics \& Astronomy, York University, 4700 Keele St, Toronto,
ON M3J 1P3, Canada \\
$^d$PRISMA Cluster of Excellence, Institut f\"ur Kernphysik and
Helmholtz~Institut~Mainz, Johannes Gutenberg-Universit\"at Mainz, D-55099 Mainz,
Germany \\
}

\abstract{
We study the thermal transition of QCD with two degenerate light
flavours by lattice simulations using $O(a)$-improved Wilson quarks. Temperature
scans are performed at a fixed value of $N_t = (aT)^{-1}=16$, where $a$ is the
lattice spacing and $T$ the temperature, at three fixed zero-temperature pion 
masses between 200 MeV and 540 MeV. In this range we find that the transition is
consistent with a broad crossover. As a probe of the restoration of chiral
symmetry, we study the static screening spectrum.  We observe a degeneracy
between the transverse isovector vector and axial-vector channels starting from
the transition temperature. Particularly striking is the strong reduction of the
splitting between isovector scalar and pseudoscalar screening masses around the
chiral phase transition by at least a factor of three compared to its value at
zero temperature. In fact, the splitting is consistent with zero within our
uncertainties. This disfavours a chiral phase transition in the $O(4)$
universality class.
}

\begin{document}

\maketitle

\section{Introduction}
\label{sec:intro}

Nuclear matter under extreme conditions of high temperatures $T$ and/or baryon
chemical potential $\mu_B$ is the subject of intense experimental and
theoretical studies in nuclear, particle and astro-physics. One of the salient
features of strongly interacting matter is the high-temperature transition from
the hadronic phase to the deconfined quark-gluon
plasma (QGP). The transition takes place in a temperature regime between 100
and 300 MeV, where the QCD running coupling is strong.  Thus, a
non-perturbative investigation of the transition and the properties of the QGP
is necessary and a lot of effort has been invested by the lattice community in the
past decades (for recent reviews
see~\cite{Brambilla:2014jmp,Szabo:2014iqa,Ding:2015ona,Bazavov:2015rfa,
Meyer:2015wax}).

\begin{figure}[t]
\centering
\includegraphics{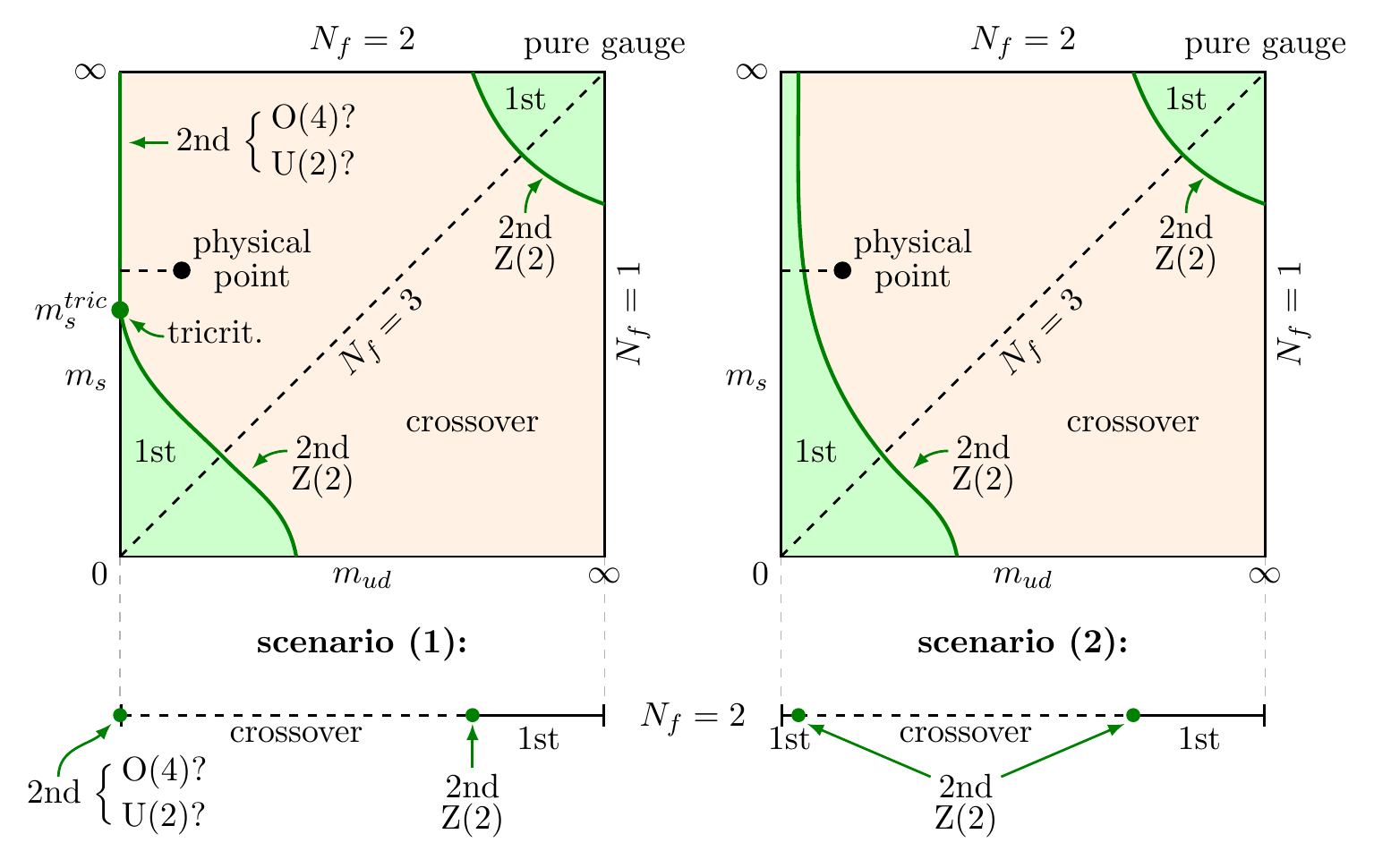}
\caption{The two possible scenarios for the quark mass dependence of the phase
structure of QCD at zero chemical potential. The lower lines highlight the
dependence of the phase structure in the $N_f=2$ case of the $u$ and $d$ quark
masses.}
\label{fig:columbia}
\end{figure}

Because lattice studies of the QCD transition at finite baryon chemical potentials 
are severely hampered by the sign problem, the QCD phase diagram remains largely unknown.
Even at zero baryon density, the nature of the 
thermal transition with light quark masses approaching the chiral limit is not yet determined in the continuum.
Knowledge of this important limit would also help to
constrain the phase diagram at non-zero $\mu_B$. Fig.~\ref{fig:columbia}
summarises the current knowledge about the order of the thermal transition
for vanishing baryon density in the
$(m_{ud},m_s)$-plane, where $m_{ud}$ is the mass of the degenerate up and
down quarks and $m_s$ the strange quark mass. In the opposite limits of pure
gauge theory and QCD with three  massless quarks, there are true first-order phase
transitions associated with the breaking of centre symmetry \cite{Yaffe:1982qf},
and the restoration of the SU(3) chiral symmetry \cite{Pisarski:1983ms},
respectively. These get weakened by the explicit breaking of those symmetries
by finite fermion masses, until they disappear along second order critical lines. 
For intermediate quark masses, the finite
temperature transition is then merely an analytic
crossover.

There is plenty of evidence that the physical quark mass configuration realised in nature
is in the crossover 
region. Early results based on the staggered fermion discretisation \cite{Aoki:2006we,Bazavov:2011nk}
have been confirmed by domain wall fermions \cite{Bazavov:2012qja,Bhattacharya:2014ara}
and simulations with Wilson
fermions are approaching the physical point as
well~\cite{Umeda:2012er,Borsanyi:2012uq,Borsanyi:2015waa,Umeda:2015jjs}.
The critical line separating the first order chiral transitions from the crossover region, the 
chiral critical line, has been mapped out on coarse lattices and 
is in the $Z(2)$ universality class of the 3d Ising model
\cite{Karsch:2001nf,deForcrand:2006pv}. 
The critical line in the heavy quark region, the deconfinement critical line,  is in the same universality 
class and was mapped out 
on coarse lattices simulating a hopping expanded determinant~\cite{Saito:2011fs} and a 3d effective
lattice theory~\cite{Fromm:2011qi}.
However, the location of the critical lines in the quark mass plane is subject
to severe cut-off effects.
With standard staggered fermions the $N_f$ = 3 critical pion mass is at around
two to three times the physical quark mass
\cite{Karsch:2001nf,deForcrand:2006pv},
yet one finds that it shrinks to nearly half that value when going from $N_t=4$
to $N_t=6$ \cite{deForcrand:2007rq}. Improved staggered
fermions can only give an upper bound for the critical mass which is around 0.1 times the physical
quark mass~\cite{Endrodi:2007gc,Ding:2011du}. First results with Wilson fermions
on the other hand see the $N_f=3$ critical endpoint at around five times the physical
quark mass \cite{Jin:2014hea}. While the latter result will still change when
going towards finer lattices, it highlights the importance of
taking the continuum limit before discussing critical behaviour.

While all current results indicate that
the critical line passes the physical point to the left, its detailed continuation 
is still largely unknown~\cite{Szabo:2014iqa}. There are two
possible scenarios~\cite{Pisarski:1983ms,Butti:2003nu,Pelissetto:2013hqa}: in
scenario (1), depicted in the left panel of Fig.~\ref{fig:columbia}, the chiral
critical line reaches the $m_{ud} = 0$ axis at some tri-critical point $m_s =
m_s^{tric}$, implying a second order transition for $N_f=2$. 
In the alternative scenario (2) the chiral critical line never
reaches the $m_{ud}=0$ axis, so that the chiral transition at $m_{ud}=0$
is first order for all values of the strange quark mass.

Many past studies have investigated the nature of the $N_f=2$ transition using
staggered~\cite{Karsch:1993tv,Karsch:1994hm,Aoki:1998wg,Bernard:1999xx,
DElia:2005bv,Cossu:2007mn,Bonati:2009yg}, O($a$)
improved~\cite{AliKhan:2000iz,Bornyakov:2009qh,Bornyakov:2011yb} or twisted
mass~\cite{Burger:2011zc} Wilson fermions and domain wall and overlap
fermions~\cite{Chiu:2013wwa,Cossu:2013uua,Tomiya:2014mma,Cossu:2015lnb},
without being able to provide a conclusive answer in the
chiral limit. The main problem in investigating scaling properties is the
similarity of the critical exponents of the universality classes in question
(cf. section~\ref{sec:scaling-theo}). A new method to exploit the known tricritical 
scaling when coming from the plane of imaginary chemical potential has been proposed
in~\cite{deForcrand:2010he}. First results on coarse lattices 
with staggered~\cite{Bonati:2014kpa} and Wilson fermions \cite{Philipsen:2015eya} agree 
on scenario (2), but show enormous quantitative discrepancies.
A similar approach proposed recently is to look at the
extension of the chiral critical line in the plane with $N_F$
additional degenerate heavy quarks~\cite{Ejiri:2012rr,Ejiri:2015vip}.

Which of these scenarios is realised depends crucially on the strength
of the anomalous breaking of the $\Ua$ symmetry at the critical temperature
in the chiral limit when the mass of the strange quark is sent to infinity, i.e.\
in the $N_f=2$ case. If the breaking of $\Ua$ remains strong, the transition will
be of second order in the $O(4)$ universality class~\cite{Pisarski:1983ms}, so
that scenario (1) will be realised. If, on the other hand, the symmetry is
`sufficiently' restored (see the discussion below about different restoration criteria), 
either scenario is possible~\cite{Butti:2003nu,Pelissetto:2013hqa}.
For scenario (1) the breaking would then likely be in
the $\U(2)\times \U(2)\to \U(2)$
universality class~\cite{Butti:2003nu,Pelissetto:2013hqa} (another symmetry
breaking pattern/universality class of the form $\SU(2)\times\SU(2)\times
Z_4\to\SU(2)$ has also been proposed~\cite{Aoki:2013zfa,Aok:2016gxw})
and the $O(4)$ universality class would be disfavoured. To be able to
distinguish between the different scenarios by looking at the
$\Ua$ symmetry, it is thus crucial to have a measure for the strength of the
breaking.

First, we recall that the $\Ua$ classical symmetry does not imply the existence
of a conserved current: the divergence of the singlet axial current
$A_\mu^0(x)$ is proportional to the gluonic operator $G\tilde G =
\epsilon_{\mu\nu\rho\sigma}G_{\mu\nu}G_{\rho\sigma}$ in the chiral limit, an
equality valid in any on-shell correlation function. For instance, the static
correlator $({\partial^2}/{\partial x_3^2})\int dx_0 dx_1 dx_2 \langle A_3^0(x)
A_3^0(0)\rangle $ is proportional to the corresponding static two-point function
of $G\tilde G$, which is certainly non-vanishing at any finite temperature. By
contrast, the corresponding two-point function of the non-singlet axial current
$A_3^a(x)$ vanishes in the chiral limit. In this particular sense, the $\Ua$
symmetry is never restored.

In this paper we use static correlators and the associated screening spectrum to
probe the $\Ua$ effects. A thermal state with an exactly restored $\Ua$ symmetry
implies that the correlators $\langle\bar\psi(x) \tau^a\psi(x) \bar\psi(0)
\tau^a\psi(0)\rangle$ and $\langle\bar\psi(x) \gamma_5\tau^a\psi(x)\;
\bar\psi(0) \gamma_5\tau^a\psi(0)\rangle$ are equal in the massless theory.
However, we expect the restoration of $\Ua$ symmetry in this sense to only be
partial, improving as the temperature is increased (see e.g.\
\cite{Dunne:2010gd}).

In the literature, the restoration of $\Ua$ symmetry has also been
discussed in a different, more restrictive sense. The observables considered
are e.g.\ the correlation functions mentioned in the previous paragraph
projected onto zero-momentum ($\int d^4x$). Similar to the
Banks-Casher relation for the chiral condensate, the difference of the
zero-momentum correlators can be expressed in terms of the spectral
density of the Dirac operator alone~\cite{Cohen:1996ng,Cohen:1997hz}.
For a density $\rho(\lambda)\sim |\lambda|^\alpha$ with $\alpha>1$, the
difference vanishes exactly in the chiral limit. In such a scenario, it is said
that $\Ua$ symmetry is restored. More generally, if the $\SU(2)$ symmetry is
restored at $\Tc$ (which we will find to be the case later), the effective
restoration of the $\Ua$ symmetry is indicated by the degeneracy of correlation
functions belonging to a $\U(2)\times\U(2)$ multiplet
(e.g.~\cite{Cohen:1996ng}). Using the restoration of the $\SU(2)$ symmetry, it
was shown by Cohen~\cite{Cohen:1996ng,Cohen:1997hz} that the degeneracy of
zero-momentum correlation functions in the multiplets is directly linked to the
eigenvalue density $\rho(\lambda)$ in the vicinity of zero. In particular, if
the spectrum of the Dirac operator develops a gap around $\lambda=0$ at $\Tc$,
the $\Ua$ symmetry becomes effectively restored. Using QCD inequalities, it has
been argued that the $\Ua$ symmetry is expected to be effectively restored as
soon as the $\SU(2)$ symmetry is intact~\cite{Cohen:1996ng}. Later it was noted
that the result using the inequalities was incorrect since the contributions
from sectors with non-zero topology had not been taken into account
properly~\cite{Evans:1996wf,Lee:1996zy}. However, the non-zero topology sectors
only contribute away from the thermodynamic limit. Only a bit later it was
shown that the eigenvalue density in the chiral limit behaves like
$\rho(\lambda) \sim \vert \lambda \vert^\alpha$ with
$\alpha>1$~\cite{Cohen:1997hz}. Using mild assumptions and Ward-Takahashi
identities for higher order susceptibilities in the framework of overlap
fermions on the lattice, it has recently been shown that in fact
$\alpha>2$~\cite{Aoki:2012yj}, meaning that not only the eigenvalue density, but
also its first and second derivative vanish at the origin, speaking strongly in
favour of a restoration of the $\Ua$ symmetry.

The relation between degeneracy of correlators and the behaviour of the low modes
of the Dirac operator triggered a number of numerical studies
of the eigenvalue spectrum in the vicinity of
$\Tc$~\cite{Bazavov:2012qja,Cossu:2013uua,Chiu:2013wwa,
Tomiya:2014mma,Dick:2015twa,Cossu:2015lnb}. Some
groups~\cite{Cossu:2013uua,Chiu:2013wwa} see a restoration of $\Ua$ at $\Tc$ in
the chiral limit (in particular, in~\cite{Chiu:2013wwa} the behaviour of
$\rho(\lambda) \sim \vert \lambda \vert^3$ was observed), while others claim
that $\Ua$ will still be broken
in the chiral limit~\cite{Bazavov:2012qja,Dick:2015twa}. The
difference between these studies is (apart from lattice spacings and volumes)
the fermion action in use, in particular, how well the actions preserve chiral
symmetry. In fact~\cite{Aoki:2012yj}, to be able to show that the breaking of
$\Ua$ still affects zero-momentum correlation functions via the low eigenvalues
of the Dirac operator it is mandatory to fulfill the requirements of: (a)
restored chiral symmetry on the lattice; (b) extrapolation to the infinite
volume limit; (c) extrapolation to the chiral limit. In particular, condition
(a) appears to play a crucial role. The reason is that any explicit breaking of
the symmetry in the chiral limit by the lattice interferes with the effective
restoration. Furthermore, it is mandatory to use the same fermion action also
for the sea quarks, as shown in~\cite{Tomiya:2014mma,Cossu:2015lnb}. Here the
authors looked at the small eigenvalues with domain-wall fermions and overlap
fermions on the domain-wall ensembles and observed a broken $\Ua$ symmetry,
actually made worse by the use of ``quenched'' overlap quarks. Only after
reweighting of the configurations to the overlap ensemble an actual restoration
of $\Ua$ in the chiral limit was observed. A possible explanation for this
effect is that the ``quenching'' of the overlap operator leads to the appearance
of non-physical near zero modes in the overlap spectrum, much like the
appearance of exceptional configurations in quenched QCD. Indeed, the cases
where a residual breaking was observed were those  where either or both, valence
and sea quarks, might not have a fully restored chiral symmetry on the lattice.
Similar conclusions have been found using chiral susceptibilities, which can
also be related to the eigenvalue spectrum,
e.g.~\cite{Bazavov:2012qja,Buchoff:2013nra,Cossu:2015lnb}. However, when
computed on the lattice, the susceptibilities suffer from contact terms, which
need to be carefully subtracted to obtain conclusive results.

In this article we present a study of the phase transition in two-flavour QCD
using non-perturbatively O($a$) improved Wilson
fermions~\cite{Sheikholeslami:1985ij} and the Wilson plaquette
action~\cite{Wilson:1974sk}. We work with a large temporal lattice extent of
$N_t=16$ throughout, which at the chiral transition
corresponds to a lattice spacing $a\approx 0.07$ to 0.08 fm.
Our pion masses range from about $ 200$ to $500\;$MeV.  In particular, we study
the pseudo-critical temperatures defined by the change in the 
Polyakov loop and the chiral condensate, pertaining to deconfinement and chiral
symmetry restoration, respectively, and check for the associated scaling in the
approach to the chiral limit. As already discussed in~\cite{Brandt:2013mba},
such a scaling analysis is not sufficient to distinguish 
between the universality classes in question. We thus direct our attention to
the strength of the $\Ua$ breaking by investigating the degeneracy pattern of
screening masses. This is complementary to other studies of the $\Ua$ symmetry
in the literature described above,
e.g. \cite{Bazavov:2012qja,Cossu:2013uua,Bhattacharya:2014ara,Tomiya:2014mma,
Cossu:2015lnb}, which are based on the eigenvalue structure of the Dirac
operator. The screening masses probe the long-distance properties the
correlators and are free of contaminations from contact terms, unlike chiral
susceptibilities. We propose a measure for the strength of the $\Ua$-breaking in
the vicinity of $\Tc$ and extrapolate it to the chiral limit. There we find it
to be consistent with zero and 3 standard deviations away from its non-zero
value at zero temperature. This suggests an effective restoration of the $\Ua$
symmetry around the critical temperature and thus a strengthening of the chiral
transition for the lattice spacing considered.

As discussed above, at finite lattice spacing an exact chiral symmetry 
is mandatory in order to study the eigenvalue spectrum of the Dirac operator
reliably. In this study we use O($a$) improved Wilson fermions.  While the
action breaks chiral symmetry at finite lattice spacing, the static screening
masses that we study approach their continuum limit with O($a^2$) corrections.
Therefore, as long as we work in a regime where these corrections are small
compared to the physical mass splittings induced by the $\Ua$ anomaly, we
should obtain qualitatively correct conclusions. If, at a given lattice spacing,
a $\Ua$-breaking mass splitting turns out to be small, a continuum extrapolation
is required to determine how small exactly the splitting is.
 
Parts of our results have already been presented at
conferences~\cite{Brandt:2010uw,Brandt:2010bn,Brandt:2012sk,Brandt:2013mba} and
were used to investigate the properties of the
pion quasiparticle in the vicinity of the
transition~\cite{Brandt:2014qqa,Brandt:2014sna,Brandt:2015sxa,Brandt:2015eka}.

The article is organised as follows: In the next section we introduce our
observables, the details of our simulations and discuss the renormalisation
and scale-setting procedures. In section~\ref{sec:results}, we present the
numerical results. We first discuss the extraction of the pseudo-critical
temperatures in section~\ref{sec:critical-temp} and try to compare the results
to the scaling predictions in the approach to the chiral limit. We also compare
our results with those from different
fermion discretisations in the literature. In 
section~\ref{sec:screning-masses} we discuss the screening masses in
the different channels, before we come to the investigation of the strength of
the breaking in the chiral limit in section~\ref{sec:ua1-breaking}. Finally we
present our conclusions in section~\ref{sec:concl}. Detailed tables collecting 
simulations parameters and results can be found in the appendices.

\section{Lattice simulations}

\subsection{Simulation and scan setup}
\label{sec:sim-setup}

Our simulations are performed using two flavours of non-perturbatively $O(a)$
improved Wilson fermions~\cite{Sheikholeslami:1985ij} and the unimproved Wilson
plaquette action~\cite{Wilson:1974sk}. We use the clover coefficient 
determined non-perturbatively in Ref.~\cite{Jansen:1998mx}. The simulations are done
employing deflation accelerated versions of the
Schwarz~\cite{Luscher:2005rx,Luscher:2007es} (DD) and
mass~\cite{Hasenbusch:2001ne} (MP) preconditioned algorithms, the latter in the
implementation of Ref.~\cite{Marinkovic:2010eg}. Both algorithms make use of the
Schwarz preconditioned and deflation accelerated solver introduced
in~\cite{Luscher:2003qa,Luscher:2007se}. As discussed in detail
appendix~\ref{app:sim-algos}, the algorithms offer a significant speedup for
large volumes and low quark masses, but also pose constraints on the available
lattice sizes.

In general there are two known procedures to vary the temperature $T=1/(\Nt
a)$. The first option is to vary the temporal extent whilst keeping the lattice
spacing fixed. The advantage of this procedure is that all physical parameters
and renormalisation constants remain fixed, making it the optimal tool for
spectroscopy (see~\cite{Brandt:2015aqk}, for instance). The disadvantage of this
procedure is that the resolution around $\Tc$ is limited, made even worse by the
use of improved algorithms (cf. appendix~\ref{app:sim-algos}). The second
option, used in this study, is to vary the lattice spacing $a$ by varying the
coupling $\beta=6/g_0^2$, known as $\beta$-scans. This offers the possibility to
obtain a fine resolution around $\Tc$, but requires a good tuning of the bare
quark mass to scan along lines of constant physics (LCPs), as well as the interpolation
of quantities needed for scale setting and renormalisation. This is particularly
demanding for Wilson fermions due to the additive quark mass renormalisation.

We will use throughout a comparatively large temporal extent of $\Nt=16$ for two reasons.
First, Wilson fermions break chiral symmetry explicitly at finite lattice
spacing. Being as close as possible to the continuum helps to reduce the
resulting effects as much as possible. Second, for our choice of lattice action
the non-perturbative determination of the improvement coefficient $\csw$ in the
two-flavour theory extends only down to $\beta=5.2$~\cite{Fritzsch:2012wq}.
Since $a\approx0.08$ fm at $\beta=5.2$ this means that $\Nt=16$ is necessary to
allow for scans in the desired temperature range.

\subsection{Scale setting and lines of constant physics}
\label{sec:lcps}

To convert our results to physical units we use the Sommer
scale~\cite{Sommer:1993ce} $r_0$ with the interpolation of the CLS results
from~\cite{Fritzsch:2012wq} discussed in appendix~\ref{app:T0-interpol}. To
convert to physical units we use the continuum result
$r_0=0.503(10)$~fm~\cite{Fritzsch:2012wq}. The temperature scans are done along
LCPs, for which we estimate the values for the bare parameter $\kappa$
corresponding to a particular quark mass by inverting the analytic relation
$m_{ud}(\beta,\kappa)$ discussed in appendix~\ref{app:lcps}. 
We test the validity of this relation a posteriori by computing $m_{ud}$ 
along the $\beta$-scan. Conventionally, quark masses will be quoted in the
$\msbar$ scheme at a renormalisation scale of 2~GeV.

Whenever we quote pion masses for our temperature scans, we imply that these are
zero temperature pion masses which correspond to the quark masses of the
respective ensemble. We estimate the pion masses from our results for $m_{ud}$
using chiral perturbation theory to next-to-next-to leading order as given
in~\cite{Bijnens:1998fm}. For this we use the low-energy constants
from~\cite{Brandt:2013dua} obtained by the fit denoted as `NNLO
$F_\pi,\,m_\pi^2$' with a mass cut of 560~MeV. The associated low-energy
constants are given in table 6 of~\cite{Brandt:2013dua}. This procedure serves the
purpose of enabling comparisons with the literature and should not be taken as a precision
computation of the zero temperature pion mass.

\subsection{Observables and renormalisation}

\subsubsection{Deconfinement and the Polyakov loop}

To investigate the deconfinement properties of the transition we look at the
associated order parameter, the Polyakov loop
\be
\label{eq:poly-loop}
L = \frac{1}{V} \sum_{\vec{x}} {\rm Tr}\Big\{ \prod_{n_0=1}^{N_t}
U_0(n_0\,a,\vec{x})\Big\} \,,
\ee
a nonzero value of which signals the spontaneous breaking of centre symmetry.
Dynamical fermions explicitly break the centre symmetry of the gauge action and
favour the centre sector of the Polyakov loop on the real axis, so that it is
sufficient to look at $\left< \Re(L) \right>$. In~\cite{Wittig:1989av} it was
found that the use of smeared links can enhance the signal in investigations of
phase transitions. We have thus also computed the real part of the
APE-smeared~\cite{Albanese:1987ds} Polyakov loop, $\left< \Re(L)_S \right>$,
using 5 steps of APE smearing with a parameter of 0.5 multiplying the staples.
The Polyakov loop susceptibility is given by
\be
\label{eq:poly-susc}
\chi_L = V \left( \left< \Re(L)^2 \right> - \left< \Re(L) \right>^2
\right)
\ee
and similarly for the smeared Polyakov loop. In order to have a quantity with a
well defined continuum limit the Polyakov loop requires multiplicative
renormalisation~\cite{Kaczmarek:2002mc}. Here we will ignore this issue and work
with the unrenormalised Polyakov loop. Since the renormalisation factor is expected 
to behave monotonically at the $\beta$ values corresponding to the critical
region, we do not expect the typical S-shape of the Polyakov loop vs.\
temperature graph to be affected.

\subsubsection{The chiral condensate}
\label{sec:chcond}

A second aspect of the transition to the quark gluon plasma is the restoration of
chiral symmetry. In the chiral limit, the associated order parameter  is the
chiral condensate
\be
\label{eq:chicond}
\cond = - \frac{T}{V} \frac{\partial \ln(Z)}{\partial m_{ud}}\,.
\ee
It governs the response of the system with respect to the external `field' which
breaks the symmetry explicitly, i.e., the quark mass $m_{ud}$. The bare chiral
condensate is given by
\be
\label{eq:chicond-lat}
 \cond\bare = - \frac{N_f T}{V} \big\langle \Tr\big(D^{-1}\big) \big\rangle_{U}
\,,
\ee
where $D$ is the Dirac operator and the expectation value on the right-hand side
is taken with respect to the gauge field.

The associated susceptibility
\be
\label{eq:chisusc}
\chi_\cond = \frac{T}{V} \frac{\partial^2 \ln(Z)}{\partial m_{ud}^2}
\ee
consists of a disconnected (the terms in the curly brackets) and a connected
part,
\begin{eqnarray}
\label{eq:chisusc-lat}
\chi_\cond\bare & = & \frac{T N_f}{V} \left[ \left\{ \left<
\Tr\big(D^{-1}\big)^2 \right>_U  - \left< \Tr\big(D^{-1}\big) \right>^2_U \right\} -
\frac{1}{2} \left< \Tr\big(D^{-1}D^{-1}\big) \right>_U \right] \nonumber \\
 & = & \chi_\cond\bare\vert_{\rm disc} -  \frac{T N_f}{2V} \left<
\Tr\big(D^{-1}D^{-1}\big) \right>_U \,.
\end{eqnarray}
In the region around $T_c$ the disconnected part has been found to
dominate the transition signal in the
susceptibility~\cite{Bazavov:2011nk,DElia:2005bv,Aarts:2014nba}.
Close to the chiral limit, however, this statement does not
necessarily hold. The connected part only receives contributions from
isovector states, while the disconnected part receives contributions
both from isovector and isoscalar states.
Since an unbroken $U_A(1)$ symmetry would imply light isovector scalar
states, the relevant magnitude of the two contributions is an
important indicator of the nature of the transition. Here we will
focus on the disconnected part of the susceptibility and leave the
comparison between the connected and the disconnected susceptibility
for future publications.

Due to mixings with operators of lower dimension, the condensate 
contains cubic, quadratic and linear divergences, and therefore requires
additive renormalisation~\cite{Karsten:1980wd,Bochicchio:1985xa}. In addition, 
it renormalises multiplicative with the renormalisation factor associated with
the scalar density, $Z_S$, which is equivalent to the inverse of the mass
renormalisation factor for Wilson fermions, $Z_S=Z^{-1}_m$~\cite{Giusti:1998wy},
where $Z_m$ is defined in appendix~\ref{app:lcps}.
Neither the additive nor the multiplicative renormalisation depends on the
temperature, so that we can cancel the divergent terms by subtracting the chiral
condensate at $T=0$. The associated difference renormalises multiplicatively,
\be
\label{eq:chicond-ren}
\cond\ren(T) = Z^{-1}_m \left[ \cond\bare(T) - \cond\bare(0) \right] \,.
\ee
For the determination of $Z_m$ we can use eqs.~\refc{eq:ren-mud-bare},
\refc{eq:ren-mud-pcac} and \refc{eq:rel-bare-pcac} to obtain
\be
\label{eq:Zm-factor}
Z_m = \frac{Z_A}{Z_P} Z_{\rm PCAC}(\beta) ( 1 + b_m \, a\bar{m} ) \,,
\ee
which is correct up to $O(a^2)$. $Z_{\rm PCAC}(\beta)$ can be taken from the
fit in appendix~\ref{app:lcps}.

Using axial Ward identities (AWIs) one can also define an observable which is
free of cubic divergences and reproduces the chiral condensate in the chiral
limit~\cite{Bochicchio:1985xa,Giusti:1998wy}. The axial Ward identity in the
form integrated over spacetime (and up to a contact term at $y=x$) is given by
\be
\label{eq:awi-cond}
\frac{1}{N_f} \left[ \cond\bare(x) - b_0 \right] = 2 m_{\rm PCAC} a^4 \sum_y
\left< P(x) P(y) \right> - Z_A a^4 \sum_y \partial_y^\mu \left< P(x) A_\mu(y)
\right> \,,
\ee
where $m_{\rm PCAC}$ is the bare PCAC quark mass and $b_0$ represents the cubic
divergence in the bare condensate. At finite quark mass the second
term on the r.h.s. vanishes due to the absence of a true Goldstone
boson~\cite{Giusti:1998wy} and we can use the first term as the definition of a
bare subtracted condensate
\be
\label{eq:chicond-sub}
\cond\bare_{\rm sub} = 2 N_f m_{\rm PCAC} \left<\overline{PP}\right>
\ee
with
\be
\label{eq:pp-term}
\overline{PP} = \frac{a^4}{N_t N_s^3} \sum_{x,y} P(x) P(y) = 
\frac{T}{V} \Tr\big( D^{-1} \gamma_5 D^{-1} \gamma_5 \big) \,.
\ee
$\cond\bare_{\rm sub}$ still suffers from additive quadratic and linear
divergences and renormalises multiplicatively with
$Z_P$~\cite{Giusti:1998wy}. Again we can subtract the residual additive
divergences using the $T=0$ counterpart, so that we obtain an alternative
renormalised chiral condensate
\be
\label{eq:chicond-sub-ren}
\cond\ren_{\rm sub}(T) = Z_P \left[ \cond\bare_{\rm sub}(T) -
\cond\bare_{\rm sub}(0) \right] \,.
\ee
We define the susceptibility of the quantity (\ref{eq:chicond-sub}) as
\be
\label{eq:chisusc-sub}
\bar{\chi}_{\cond_{\rm sub}}\bare = 4 N_f^2 V m^2_{\rm PCAC} \left[
\left< \overline{PP}^2 \right> - \left< \overline{PP} \right>^2 \right] \,.
\ee
Note that, while $\bar{\chi}_{\cond_{\rm sub}}\bare$ is not equivalent to the
disconnected chiral susceptibility in eq.~\refc{eq:chisusc-lat}, it is
expected to show a peak at the position of the chiral transition.
The subtraction of the condensates at $T=0$ requires the measurement of
$\cond\bare$ and $\left<\overline{PP}\right>$ on zero temperature ensembles.
Here we use the set of $N_f=2$ ensembles generated within the CLS effort and the
interpolation discussed in appendix~\ref{app:chicond-inter}.

\subsubsection{Mesonic correlation functions and screening masses}

Mesonic correlation functions are a valuable probe of the properties of the
QGP~\cite{DeTar:1987ar,DeTar:1987xb}. 
Let 
\be
\label{eq:meson-corrfunc}
C_{XY}(x_\mu) = \int d^3x_\perp \left< X(x_\mu,\ve{x}_\perp)Y(0) \right> \,,
\ee
be the correlation function of two operators $X$ and $Y$.
The equality of two correlation functions in channels of different quantum numbers
signals the restoration of the associated chiral symmetry. 
Here $x_\mu$ is the coordinate of the direction in which the
correlation function is evaluated and $\ve{x}_\perp$ is the coordinate vector
in the orthogonal subspace. 
The isovector correlation functions of interest for the chiral transition are
the vector ($V$) vs.\ axial-vector ($A$), and the 
pseudoscalar ($P$) vs.\ scalar ($S$) channels, related by
\be
\label{eq:av-symm}
C_{VV} \quad \stackrel{\text{SU(2)}}{\longleftrightarrow} \quad C_{AA} \qquad
\text{and} \qquad C_{PP} \quad \stackrel{\Ua}{\longleftrightarrow} \quad C_{SS}
\,.
\ee
We choose the isovector channels as observables because they are free of
disconnected diagrams, and the correlation functions can therefore be obtained
with greater accuracy. The bilinear operators for the different channels are
listed in table~\ref{tab:scm-obs}.

\begin{table}
 \begin{center}
 \begin{tabular}{l|c|c|c|c}
  \hline
  \hline
  channel & \hspace*{3mm} $S$ \hspace*{3mm} & \hspace*{3mm} $P$ \hspace*{3mm} &
\hspace*{3mm} $V$ \hspace*{3mm} & \hspace*{3mm} $A$ \hspace*{3mm} \\
  \hline
  $\Gamma$ & $\mathbf{1}$ & $\gamma_5$ & $\gamma_i$ & $\gamma_i\gamma_5$ \\
  \hline
  \hline
 \end{tabular}
 \caption{Bilinear operators $\bar{\psi}\Gamma\psi$ used for the screening correlators. Here $i=1,2$
for screening masses in $x^3$-direction.}
 \label{tab:scm-obs}
 \end{center}
\end{table}

While temporal correlation functions $C_{XY}(x_0)$ can be related to the real-time spectral
densities (see~\cite{Meyer:2011gj}), here  we are interested in spatial
correlation functions $C_{XY}(x_3)$, which are related to the screening states of the plasma. In
particular, the leading exponential decay of the correlator $C_{XX}(x_3)$  defines the 
lowest-lying `screening mass' $M_X$ associated with the quantum numbers
of the operator $X$. Screening masses can be interpreted as the inverse length
scale over which a perturbation is screened by the plasma. If a symmetry imposes the equality
of two correlation functions, it must also imply the degeneracy of the corresponding screening masses.
The latter are thus quantities sensitive to the restoration of the symmetry.
Consequently, the screening masses in the $V$ and $A$ channels provide an alternative way
of defining the chiral symmetry restoration temperature. In contrast to
susceptibilities, defined by the integrated correlation function, screening
masses probe the long-distance properties of the correlation functions
and thus do not suffer from contact terms.

Apart from their relation to chiral symmetry, mesonic screening masses are
valuable quantities to probe the medium effects of the plasma and, at high temperatures,
to test the applicability of perturbation theory. They  have been studied in lattice QCD
for a long time (for a review of early results see~\cite{Karsch:2003jg} and for
more recent studies~\cite{Cheng:2010fe,Banerjee:2011yd}). In the 
high-temperature limit, all screening masses approach the asymptotic value
$M^\infty=2\pi T$~\cite{Eletsky:1988an,Florkowski:1993bq}. The leading order
correction from the interactions has been computed in perturbation theory and is
known to be positive~\cite{Laine:2003bd,Alberico:2007yj}. Static and non-static
screening masses can also be computed within an effective theory approach and
provide an indirect probe for real-time physics in the Euclidean lattice
setup~\cite{Brandt:2014uda,Brandt:2014cka}.

\subsection{Investigating the order of the transition in the chiral limit}

\subsubsection{Critical scaling}
\label{sec:scaling-theo}

The main question driving the present study is the nature of the phase transition in the chiral limit.
Simulations with vanishing quark masses are currently impossible; in order to extract information on the order of the
transition, it is customary to investigate the scaling of various observables in the approach to the
critical point $(0,0)$ in the parameter space of reduced temperature $\tau=(T-T_c)/T_c$ and
external field $h$. The scaling laws can be derived from the scaling of the free
energy $F$ (see~\cite{Pelissetto:2000ek}). The variable playing the role of the external field depends on the
particular scenario (see section~\ref{sec:intro}). If the second order
scenario (scenario (1)) is realised, no matter whether the universality class
is O(4) or the one from the U(2)$\times$U(2)$\to$U(2) scenario~\footnote{We
will denote the U(2)$\times$U(2)$\to$U(2) scenario in short as the U(2) scenario
from now on.}, the critical point is located in the
chiral limit and the chiral condensate constitutes a true order parameter. In
this case the external field $h$ is proportional to the (renormalised) quark
mass $m_{ud}$. In the first-order scenario, depicted in the right panel of 
figure~\ref{fig:columbia}, the critical point is located at a finite quark mass
$m_{ud}^{\rm cr}$, so that chiral symmetry is broken explicitly at the critical
point. One must in general expect that the external field is given by a linear
combination of $\delta m = m_{ud}-m_{ud}^{\rm cr}$ and $\tau$.
Furthermore, $\cond$ no longer constitutes a true
order parameter. The situation is analogous to the approach of the chiral critical line
along the $N_f=3$ axis~\cite{Karsch:2001nf}.

Here we will perform an analysis based on the scaling of the order
parameter or the transition temperature $\Tc$ with the external field. In the
vicinity of the critical point a true order parameter $\Theta$ satisfies the
scaling relation (see
e.g.~\cite{AliKhan:2000iz,Pelissetto:2000ek,Ejiri:2009ac,Bazavov:2011nk})
\be
\label{eq:order-para-scaling}
\Theta \sim h^{1/\delta} f\Big(\frac{\tau}{h^{1/(\delta\beta)}}\Big) + {\rm
s.v.}
\,.
\ee
Here $f$ is a function depending on the universality class of the
transition and s.v.\ stands for scaling violations which constitute terms that
are regular in $\tau$~\cite{Pelissetto:2000ek,Ejiri:2009ac,Bazavov:2011nk}. A
number of studies have looked at the scaling of the chiral condensate in the
approach to the chiral
limit~\cite{Burger:2011zc,AliKhan:2000iz,Ejiri:2009ac,Bazavov:2011nk,
DElia:2005bv} and found consistency with O(4) scaling.

\begin{table}[t]
\begin{center}
\small
\begin{tabular}{c|cccc|c}
\hline
\hline
UC & $\nu$ & $\gamma$ & $\beta$ & $\delta$ & Ref. \\
\hline
Z(2) & 0.6301( 4) & 1.2372( 5) & 0.3265( 3) & 4.789( 2) &
\cite{Pelissetto:2000ek} \\
O(4) & 0.7479(90) & 1.477 (18) & 0.3836(46) & 4.851(22) & \cite{Kanaya:1994qe}
\\
U(2) & 0.76(10)(5) & 1.4(2)(1) & 0.42(6)(2) & 4.4(3)(1) &
\cite{Pelissetto:2013hqa} \\
\hline
\hline
\end{tabular}
\end{center}
\caption{Critical exponents for the universality classes (UC) relevant for the
chiral transition. The critical exponents of the U(2)$\times$U(2)$\to$U(2)
universality class, we have taken the results from~\cite{Pelissetto:2013hqa}
(from the $\overline{\rm MS}$ scheme). The first error is statistical while the
second quoted error denotes a systematic uncertainty arising from the scheme
dependence. The critical exponents of the U(2)$\times$U(2)$\to$U(2)
universality class have also been obtained recently using the bootstrap
method~\cite{Nakayama:2015ikw}.}
\label{tab:crit-exponents}
\end{table}

From the scaling in eq.~\refc{eq:order-para-scaling} one can also derive a
scaling law for the critical temperature as a function of the external
field~\cite{Karsch:1993tv}. The resulting scaling relation is 
\be
\label{eq:tc-scaling}
\Tc(h) = \Tc(0) \Big[ 1 + C h^{1/(\delta\beta)} \Big] + {\rm s.v.} \,,
\ee
where $C$ is an unknown constant. It must be stressed that these scaling laws
are only valid after the thermodynamic and continuum limits have been taken.

Another problem for any study of the scaling laws is the similarity of the
critical exponents in the three potentially relevant universality classes. They are
summarised in table~\ref{tab:crit-exponents}. For the different scenarios the
combination $\delta\beta$ is given by $\approx 1.56$ for Z(2), $\approx 1.86$
for O(4) and $\approx 1.85$ for U(2). Even between the Z(2) and the O(4)
scenarios the difference is so small that very accurate results are needed to be
able to distinguish between the two. Thus one cannot draw
conclusions from the agreement of lattice data with the scaling of one universality class
alone; instead one needs to demonstrate the ability to distinguish between the
scenarios.

\subsubsection{$\Ua$ symmetry restoration}
\label{sec:ua1-estimator}

The strength of the anomalous breaking of the $\Ua$ symmetry at the transition
temperature in the chiral limit is thought to be crucial for the order of the chiral 
transition~\cite{Pisarski:1983ms,Chandrasekharan:2007up}. However, this raises
the question of how to quantify the strength of the $\Ua$-breaking. As a
possible reference value we suggest the screening mass gap between the
isovector pseudoscalar and scalar channels at $T=0$,
\be
\label{eq:scm-ua1-massrat}
\Delta M_{PS} = M_P-M_S = -m_{a0} \,,
\ee
since the pion mass
vanishes in the chiral limit. Ultimately, one would like to obtain the chirally
extrapolated value of $m_{a0}$ from lattice QCD, since this would give a
result valid for the $N_f=2$ case in the range of relevant lattice spacings.
Unfortunately, the scalar correlation function in the iso-vector channel is
rather noisy, so that a reliable extraction is currently not possible. We will
discuss a phenomenological estimate for the chiral limit in
section~\ref{sec:ua1-breaking}.

We note that in the two-flavour theory, the $a_0$ meson is expected to
be stable or almost stable, since the $\bar K K$ and $\eta \pi$ decay
channels known from experiment are missing.  Indeed, in $N_f=2$ QCD only a
flavour-singlet pseudoscalar exists, sometimes called $\eta_2$, whose
nature is closer to the physical $\eta'$ meson, and whose mass has
been estimated at about 800MeV at the physical pion
mass~\cite{Jansen:2008wv}. The lightest isovector scalar state was
found to lie between the physical $a_0(980)$ and $a_0(1450)$
states~\cite{Jansen:2009hr}. The splitting between the pion and the lightest
isovector scalar state thus provides a convenient measure for the breaking
of $\Ua$.

\section{Results}
\label{sec:results}

\subsection{Ensembles and measurement setup}
\label{sec:ensembles}

\begin{table}[t]
\begin{center}
\small
\begin{tabular}{c|cc|ccccc}
\hline
\hline
scan & Lattice & Algorithm & $\kappa$/$m_{ud}$~[MeV] & $m_\pi$~[MeV] & $T$~[MeV]
& $\tau_{U_P}$~[MDU] & MDUs \\
\hline
\scan{B1}$_\kappa$ & $16\times32^3$ & DD-HMC & 0.136500 &  & $190-275$ &
$\sim20$ & $\sim20000$ \\
\hline
\scan{C1} & $16\times32^3$ & DD-HMC & $\sim17.5$ & 300 & $150-250$ & $\sim28$ &
$\sim12000$ \\
\hline
\scan{D1} & $16\times32^3$ & MP-HMC & $\sim8.7$ & 220 & $150-250$ & $\sim16$ &
$\sim12000$ \\
\hline
\hline
\end{tabular}
\end{center}
\caption{$\beta$-scans at $N_t=16$. Listed is the temperature range in MeV, the
integrated autocorrelation time of the plaquette $\tau_{U_P}$ and the number of
molecular dynamics units (MDUs) used for the analysis. For scan
\scan{B1}$_\kappa$ configurations have been saved each
200 MDUs, for scan \scan{C1} each 40 MDUs and for scan \scan{D1} each 20 MDUs.
The measurements of the Polyakov loop and the chiral condensate have been done
each 4 MDUs. The autocorrelation times and numbers of measurements quoted here
correspond to the ones at the location of the transition.}
\label{tab:scans}
\end{table}

\begin{figure}[t]
 \centering
\includegraphics[]{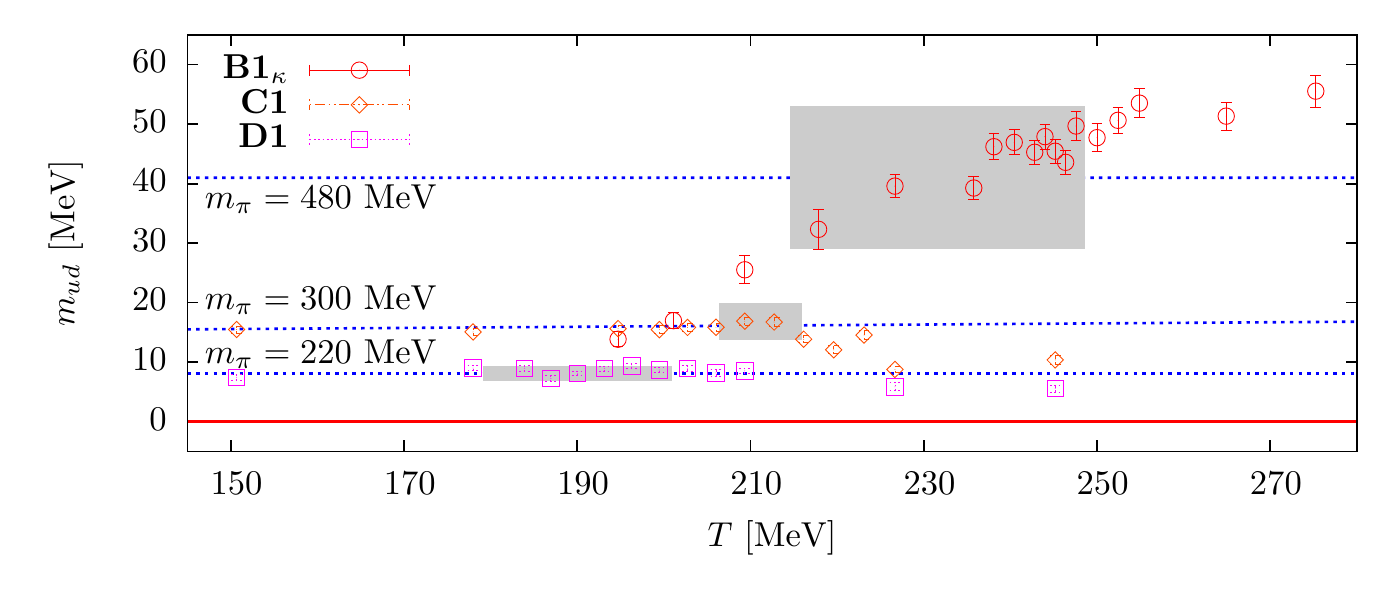}
 \caption{Simulation points in the $\{m_{ud},T\}$ parameter space. The grey
areas mark the estimates for the crossover regions.}
 \label{fig:phase-space}
\end{figure}

In this paper we present results for the chiral transition obtained from
the scans on $16\times32^3$ (and first results from $16\times 48^3$) lattices
summarised in table~\ref{tab:scans}. More
details can be found in table~\ref{tab:scan-runparams} in
appendix~\ref{app:sim-paras}.
We consider three different values of the quark mass,
corresponding to pion masses between 200 to 540~MeV. The scan corresponding to
the largest pion mass at the critical point, denoted as scan \scan{B1} (in our
naming convention the letter labels quark/pion masses while the number labels
volumes), has been done at fixed hopping parameter,
indicated by the subscript $\kappa$. Due to renormalisation and the change in
the scale, the quark mass changes with the temperature in this scan. The scans
at lighter quark masses, scans \scan{C1} and \scan{D1} with $m_\pi\approx300$
and 220~MeV, are done along LCPs. To check the tuning of the quark masses we
have measured the renormalised PCAC quark mass using the PCAC relation evaluated
in the $x_3\equiv z$-direction (see~\cite{Brandt:2013faa}). The simulation
points in the $(T,m_{ud})$ parameter space are shown in
figure~\ref{fig:phase-space}. The plot shows that the tuning of the quark mass
works well in the region below $\Tc$, while we see that we get smaller quark
masses than expected above $\Tc$. It is unclear to us whether this is a cutoff effect
or if our interpolation just becomes worse in this region (cf. appendix~\ref{app:lcps}).

The quantities relevant for the transition temperature, i.e.\ the Polyakov loop
and the chiral condensate, have been measured during the generation of the
configurations with a separation of 4 MDUs. For the measurement of the
condensate we have used 4 hits with a $Z_2\times Z_2$ volume source. The
exception is the \scan{B1}$_\kappa$ scan, where the chiral condensate has only
been measured on the stored configurations, using 100 hits. The screening
masses have been measured on the stored configurations. For scan
\scan{B1}$_\kappa$, configurations have been saved every 200 MDUs, for scan
\scan{C1} every 40 MDUs and for scan \scan{D1} every 20 MDUs. The results for the
expectation values of Polyakov loops, the chiral condensates and screening
masses are tabulated in tables~\ref{tab:scan-results} and~\ref{tab:scm-results}
in appendix~\ref{app:sim-paras}.

\subsection{The pseudocritical temperature}
\label{sec:critical-temp}

The first step of our investigation of the thermodynamics of QCD is the
extraction of the pseudocritical temperatures. Since we are dealing with a
crossover rather
than a true phase transition there is no unique definition of the critical
temperature and estimates for $\Tc$ will depend on the defining observable. To
determine $\Tc$ we will primarily look at the Polyakov loop and the chiral
condensate. In particular, the inflection point of the Polyakov loop will
define the deconfinement transition temperature $\Tc^{\rm dc}$, while the peak
in the susceptibility of the chiral condensate defines the temperature of
chiral symmetry restoration, which will be our main estimate for $\Tc$.

\subsubsection{Polyakov loops}

\begin{figure}[t]
 \centering
\includegraphics[]{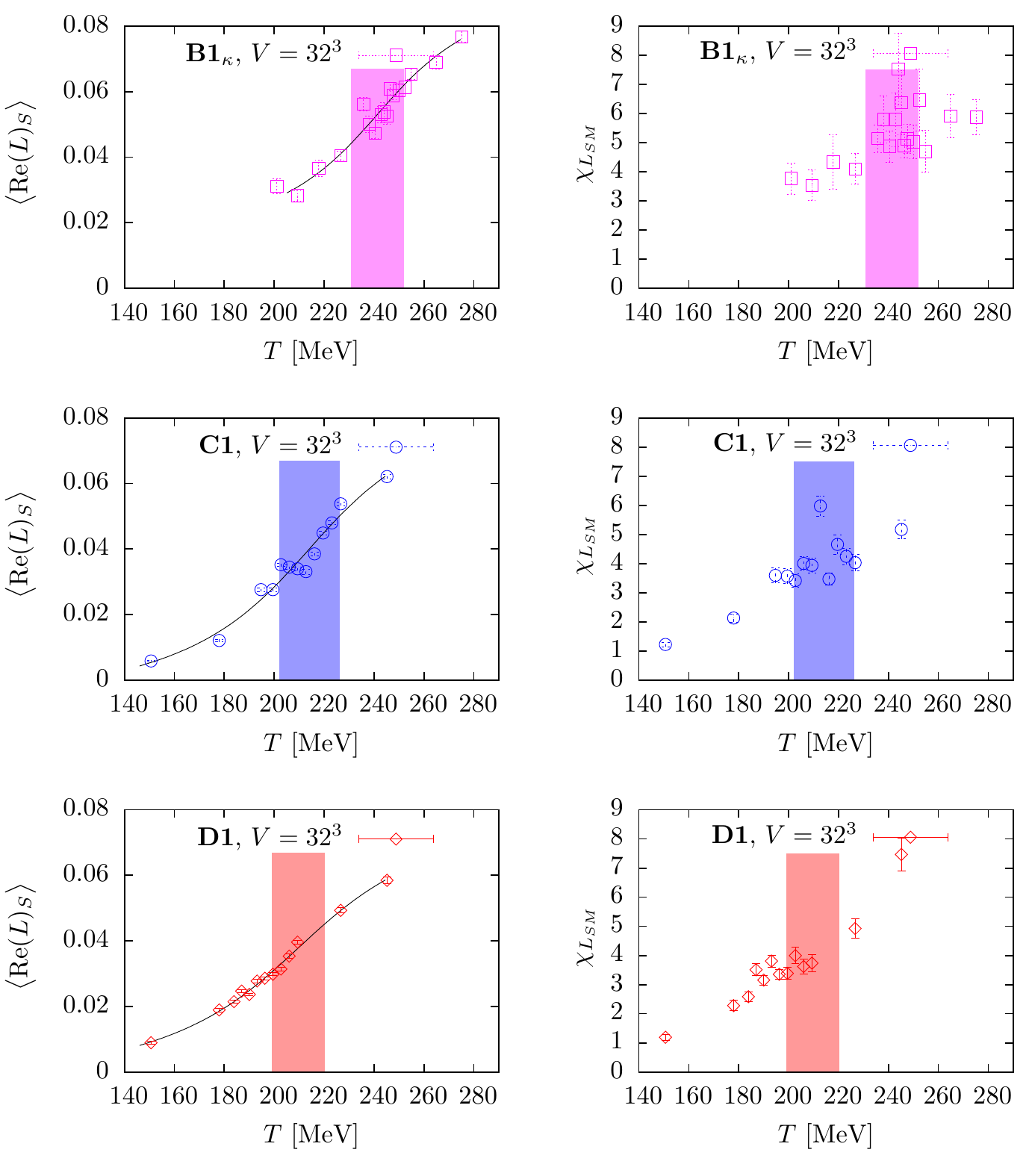}
 \caption{Results for the (unrenormalised) APE smeared Polyakov loop (left) and
its susceptibility (right) for scans \scan{B1}$_\kappa$, \scan{C1} and
\scan{D1} (from top to bottom). The shaded areas indicate the estimates for the
transition regions and the black lines are the results from the fit of the
Polyakov loop expectation value to the arctangent form.}
 \label{fig:poly-loops}
\end{figure}

We start with the extraction of $\Tc^{\rm dc}$ using the (unrenormalised)
smeared Polyakov loop. We use $\left< \Re(L)_S \right>$ since it typically shows
stronger signals for the transition. We have, however, checked the agreement
with the results for the unsmeared Polyakov loop explicitly (see
also~\cite{Brandt:2010uw}). The results are shown in
figure~\ref{fig:poly-loops}. The observable $\left< \Re(L)_S \right>$ in scans \scan{C1} and
\scan{D1} develops the S-shape characteristic for a phase transition, with some
fluctuations in the vicinity of the inflection point. For scan
\scan{B1}$_\kappa$ the S-shape is not as prominent, possibly due to the
limited temperature range explored. To extract the inflection point we have fitted
$\left< \Re(L)_S \right>$ to the form of an arctangent. To check the model
dependence of the results we have performed alternative fits using the Gaussian
error function. Both fits tend to describe the data reasonably well and give
similar $\chi^2/$dof values. The resulting curves from the arctangent fits are
shown as black lines in figure~\ref{fig:poly-loops}. The results for the
associated transition temperatures $\Tc^{\rm dc}$ are given in
table~\ref{tab:trans-points}. Evidently the estimate for the uncertainty of
the inflection point from the fit cannot be reliable due to the strong
fluctuations in its vicinity. To account for this additional uncertainty we have
assigned another systematic error of 10 MeV to the result from the fit,
reflecting the size of the interval where we observe deviations from the smooth
behaviour of the Polyakov loop. The shaded areas in figure~\ref{fig:poly-loops}
represent the estimates for the transition region. In the vicinity of $\Tc^{\rm
dc}$ the Polyakov loop susceptibility increases and shows fluctuations that can
be interpreted as the onset of peak-like behaviour. Owing to this typical
behaviour, we suspect that $\Tc^{\rm dc}$ could be somewhat overestimated for
scan \scan{D1}.

\begin{table}[t]
\begin{center}
\small
\begin{tabular}{c|cc|cc}
\hline
\hline
scan & $\Tc^{\rm dc}$ [MeV] & $\Tc$ [MeV] & $m_{ud}^C$ [MeV] & $m_\pi^C$ [MeV]
\\
\hline
\scan{B1}$_\kappa$ & 241(5)(6)(10) & 232(18)(6) & 41 (11)(2) & 485(55)(20) \\
\scan{C1}          & 214(9)(3)(10) & 211( 5)(3) & 16.8(30)(7) & 300(27)( 9) \\
\scan{D1}          & 210(6)(3)(10) & 190(10)(5) & 8.1(12)(4) & 214(14)( 8) \\
\hline
\hline
\end{tabular}
\end{center}
\caption{Results for the pseudocritical deconfinement, $\Tc^{\rm dc}$, and
chiral symmetry restoration, $\Tc$, temperatures and the associated critical
value of the quark mass with its zero-temperature pion mass pendant. The first
error reflects the uncertainty of the extraction of the pseudocritical
temperatures due to the fit, the second error accounts for scale setting (and
renormalisation). For $\Tc^{\rm dc}$ the third error is the associated
systematic error as explained in the text.}
\label{tab:trans-points}
\end{table}

\subsubsection{Chiral condensate and its susceptibility}

To estimate the chiral symmetry restoration temperatures, $\Tc$, we use the
renormalised disconnected susceptibility, of the chiral condensate without
$T=0$ subtractions. In particular, we have $Z_m^2 \chi_\cond\bare\vert_{\rm
disc}$ and $Z_P^2\bar{\chi}_{\cond_{\rm sub}}\bare$ for condensate and
subtracted condensate, respectively.
Note that the additive renormalisation discussed in
section~\ref{sec:chcond} has not been taken into account. However, the
position of the peak in the susceptibility should not be affected, since the
additive renormalisation gives regular contributions around $T_C$.

\begin{figure}[t]
 \centering
\includegraphics[]{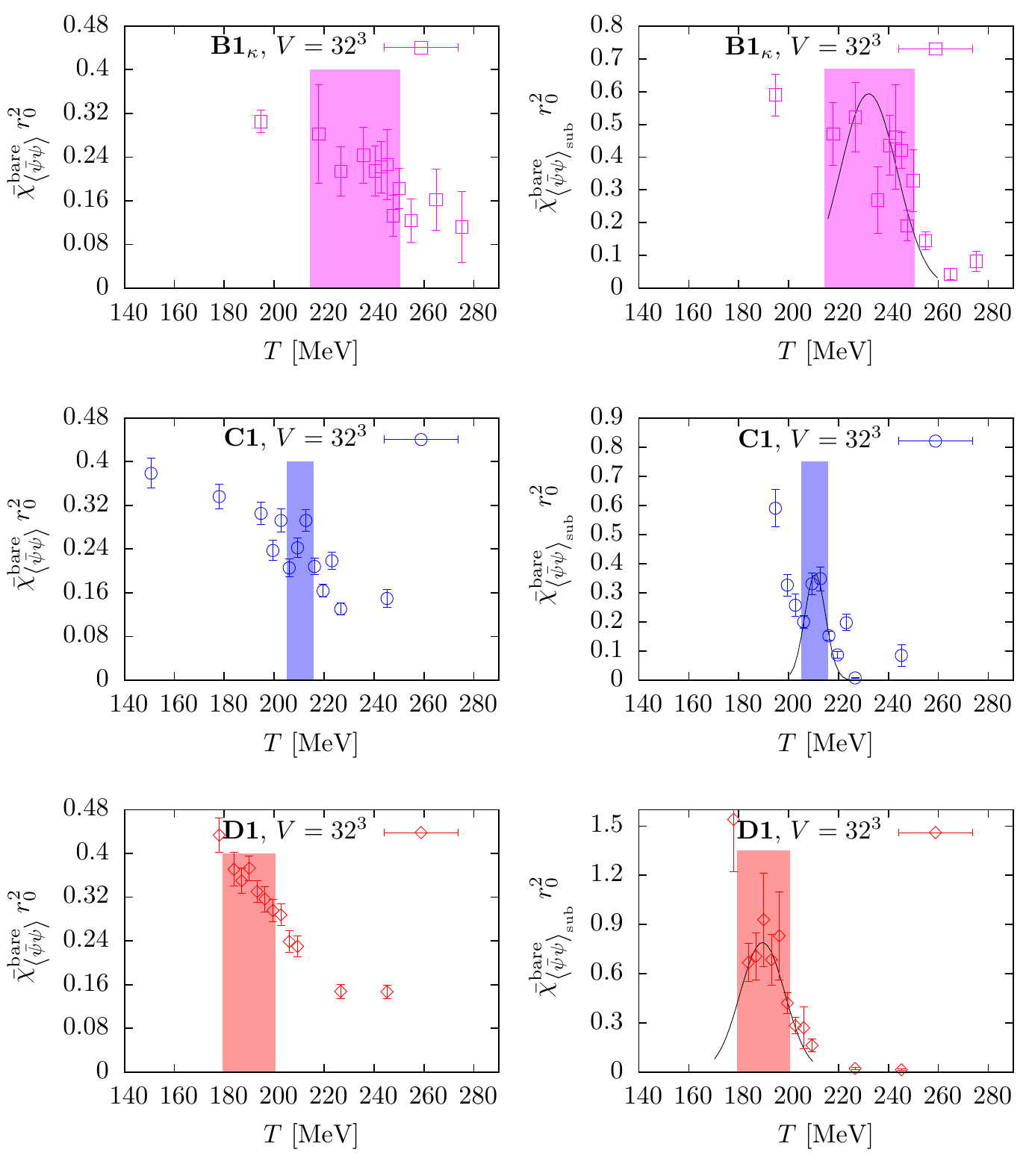}
 \caption{Results for the disconnected susceptibility of the condensate (left)
and the subtracted condensate (right) for scans \scan{B1}$_\kappa$, \scan{C1}
and \scan{D1} (from top to bottom). The shaded areas indicate the estimates for
the transition regions and the black lines are the results from the fit of the
susceptibility to a Gaussian.}
 \label{fig:susc-cond-bare}
\end{figure}

Figure~\ref{fig:susc-cond-bare} displays the results for the disconnected
susceptibilities for the unsubtracted and subtracted bare condensates. We
define the pseudocritical temperature for chiral symmetry restoration through
the position of the peak in the susceptibility of the subtracted condensate. To
determine $\Tc$ we fit the susceptibility to a Gaussian. Since the error
estimate for $\Tc$ from the fit will likely underestimate the true uncertainty
we take the full spread of points included in the fit as a conservative error
estimate. The resulting values for $\Tc$ are given in
table~\ref{tab:trans-points} and are
shown as shaded areas in figure~\ref{fig:susc-cond-bare}. The black curves
correspond to the fit. Scan \scan{B1}$_\kappa$ is a problematic case, since, due
to the change of the quark mass, the scan remains longer in the vicinity of the
critical region, $T_C(m_{ud})$. Consequently, the peak is broad and we
obtain large uncertainties for both, $\Tc$ and $m_{ud}^C$. Comparing the results
for $\Tc^{\rm dc}$ with $\Tc$, we see that they mostly agree within errors. The
exception is \scan{D1}, where we find that $\Tc^{\rm dc}$ lies somewhat above
$\Tc$. Like other studies in the literature we see that $\Tc$ decreases
with the quark
mass~\cite{Karsch:1993tv,Karsch:1994hm,Aoki:1998wg,Bernard:1999xx,
DElia:2005bv,Cossu:2007mn,Bonati:2009yg,AliKhan:2000iz,Bornyakov:2009qh,
Bornyakov:2011yb,Burger:2011zc}, which is a general feature, persisting even
when dynamical heavy quarks are included
(e.g.~\cite{Borsanyi:2015waa,Umeda:2015jjs}).

\begin{figure}[t]
\begin{center}
\includegraphics[]{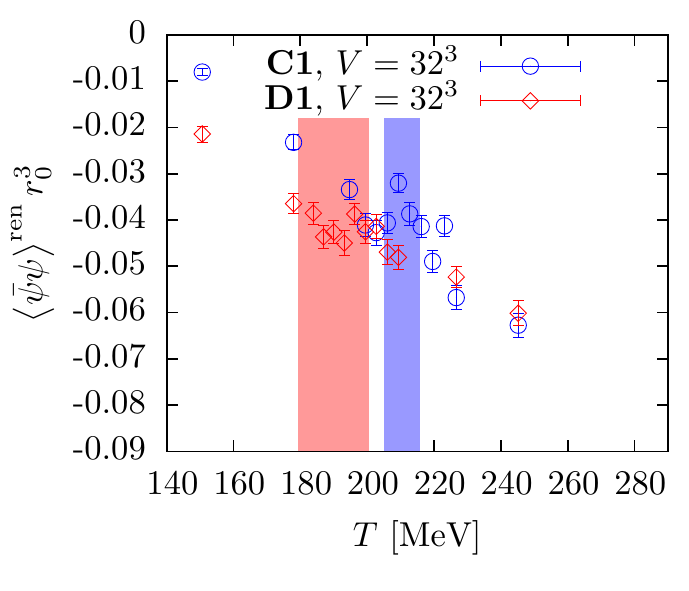} \hfil
\includegraphics[]{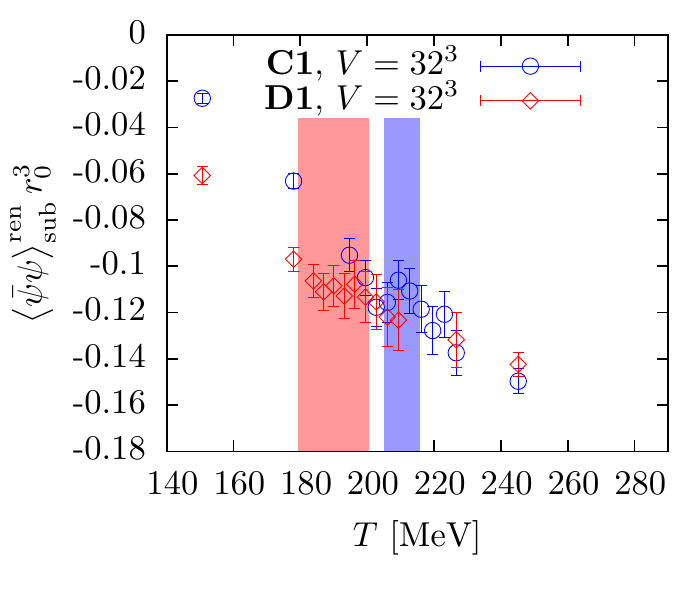}
\end{center}
\caption{Results for the renormalised standard (left) and subtracted (right)
condensate in scans \scan{C1} and \scan{D1}. The shaded areas display the
estimates for $\Tc$.}
\label{fig:cond-renorm}
\end{figure}

In figure~\ref{fig:cond-renorm} we show the results for the fully renormalised
condensate and the fully renormalised subtracted version for scans \scan{C1} and
\scan{D1}. As expected, the condensates start close to zero and show a rapid
decrease in the approach to $\Tc$. Around $\Tc$ both condensates show
fluctuations, especially the standard condensate fluctuates quite strongly.
We note that $T_C$, as defined by the inflection
point of the renormalised condensate, does not necessarily have to
agree with the peak of the susceptibility for a broad crossover.

We also have preliminary data for the condensates
from simulations on $16\times48^3$ lattices. These were mainly
used to check finite size effects on the screening masses below and
are insufficient for a precise finite size scaling analysis. However,
they are fully consistent with saturating susceptibilities and a
crossover, as expected.

\subsubsection{Scaling in the approach to the chiral limit}
\label{sec:scaling-res}

\begin{figure}[t]
\begin{center}
\includegraphics[]{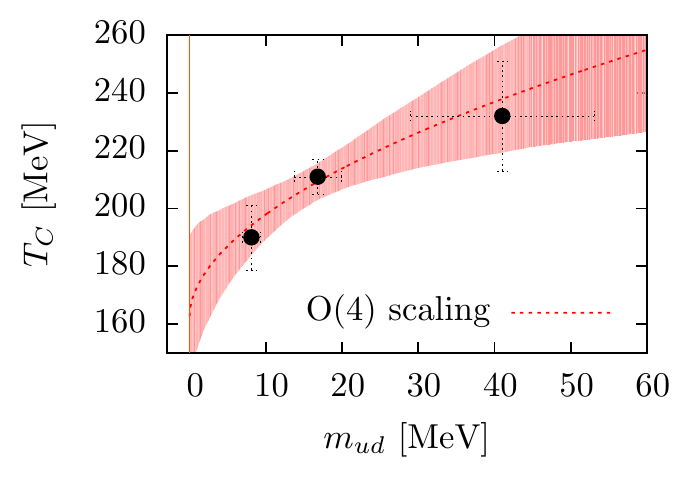} \hfil
\includegraphics[]{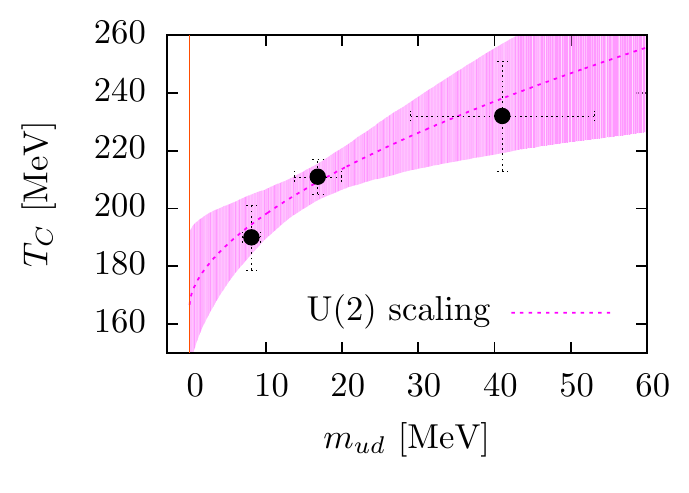}
\end{center}
\caption{Results from the scaling fits to $\Tc$ using the critical exponents
from the O(4) (left) and U(2) (right) scenarios.}
\label{fig:tc-scaling-1}
\end{figure}

Following section~\ref{sec:scaling-theo} we will now try to extract information
about the order of the transition in the chiral limit by looking at the scaling
of the temperatures with the quark mass. With three transition temperatures at
our disposal and their relatively large uncertainties we are not in the position
to extract the critical exponents from a fit of the data for $\Tc$. Instead, we
fix the critical exponents to the ones from the different universality classes
and check whether any particular scenario is favoured by our data.
We start by fitting the results for $\Tc$ from table~\ref{tab:trans-points} to
eq.~\refc{eq:tc-scaling} with the critical exponents from the O(4) and U(2)
scenarios. The results are shown in figure~\ref{fig:tc-scaling-1}. The plot
highlights the similarity of the two curves, indicating that the scaling
of the transition temperatures alone will not be sufficient to distinguish
between the two scenarios. We note that this is likely to remain true even when
the error bars are reduced by an order of magnitude. Potentially, a scaling
analysis of the order parameter and its susceptibility
(see~\cite{Bazavov:2011nk}) might help in this respect. However, this would
demand the knowledge of the scaling function from
eq.~\refc{eq:order-para-scaling} for the U(2) case. A distinction will
only be possible if the scaling functions differ significantly.

\begin{figure}[t]
\begin{center}
\includegraphics[]{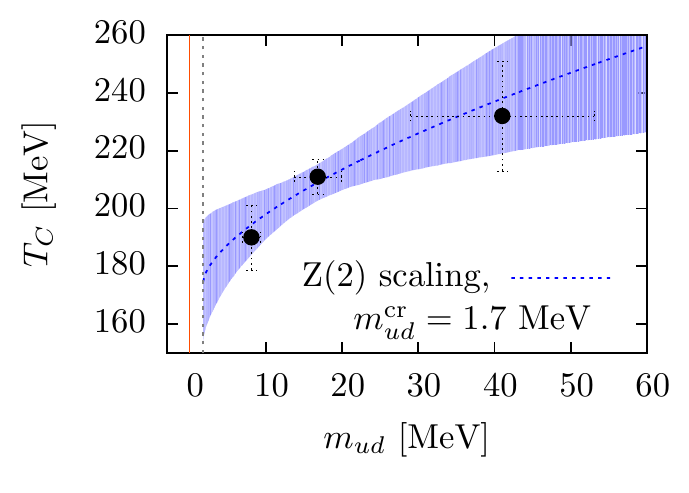}
\end{center}
\caption{Results for the scaling fit for the first order scenario,
i.e. the Z(2) universality class, with $m_{ud}^{\rm cr}=1.7$~MeV.}
\label{fig:tc-scaling-2}
\end{figure}

In the first order scenario we also have to fix the value of the critical quark
mass $m_{ud}^{\rm cr}$. Since $m_{ud}^{\rm cr}$ is poorly constrained by the
fit, we can try fits for different fixed values of $m_{ud}^{\rm cr}$
and look for minima in $\chi^2/$dof. We find that, not unexpectedly,
$\chi^2/$dof is very flat and does not exhibit a minimum. Furthermore,
$\chi^2/$dof is always of the same order as for the second order fits. As a
typical case we show the curve obtained for $m_{ud}^{\rm cr}=1.7$~MeV in
figure~\ref{fig:tc-scaling-2}. As for the scaling with the O(4) and U(2)
critical exponents, the Z(2) curve agrees very well with the data and is hardly
distinct from the curves of figure~\ref{fig:tc-scaling-1}.

Obviously, none of the scenarios is ruled out by the above analysis. These
findings are in agreement with the results from the tmfT
collaboration~\cite{Burger:2011zc}. When we assume that the data shows
consistency with one of the second order scenarios and extract the associated
critical temperatures in the chiral limit we obtain
\be
\label{eq:tc-chiral}
\left. \Tc(0) \right|_{\rm O(4)} = 163(27) \text{ MeV} \quad \text{and} \quad
\left. \Tc(0) \right|_{\rm U(2)} = 167(25) \text{ MeV} \,.
\ee
Both results are consistent with the findings from the tmfT
collaboration for O(4) scaling,
$\Tc(0)=152(26)$~MeV~\cite{Burger:2011zc}, and are on the lower side
of the results for Wilson fermions at $N_t=4$,
$\Tc(0)=171(4)$~MeV~\cite{AliKhan:2000iz}, and of the study of the
QCDSF-DIK collaboration with different $N_t$ values
$\Tc(0)=172(7)$~MeV~\cite{Bornyakov:2011yb}. Calculations using staggered
fermions only quote values for the critical coupling in the chiral
limit, without providing results for the lattice spacing. The two
results in eq.~\refc{eq:tc-chiral} indicate that the result for $\Tc(0)$ is not
sensitive to the universality class used for the extrapolation. This property is
just another manifestation of the difficulty to distinguish between the two
scenarios and shows that even a reduction of the error bars by an
order of magnitude, in combination with results at much smaller quark
masses, might not be sufficient using the scaling of the transition temperatures
alone.

\subsection{Screening masses and chiral symmetry restoration pattern}
\label{sec:screning-masses}

We now turn to the investigation of screening masses and the chiral
symmetry restoration pattern. Since the behaviour of screening masses below and
close to $\Tc$ in general depends on the quark mass we will focus on the
scans along LCPs in this section, i.e.\ on scans \scan{C1} and \scan{D1}.

\begin{figure}[t]
 \centering
\includegraphics[]{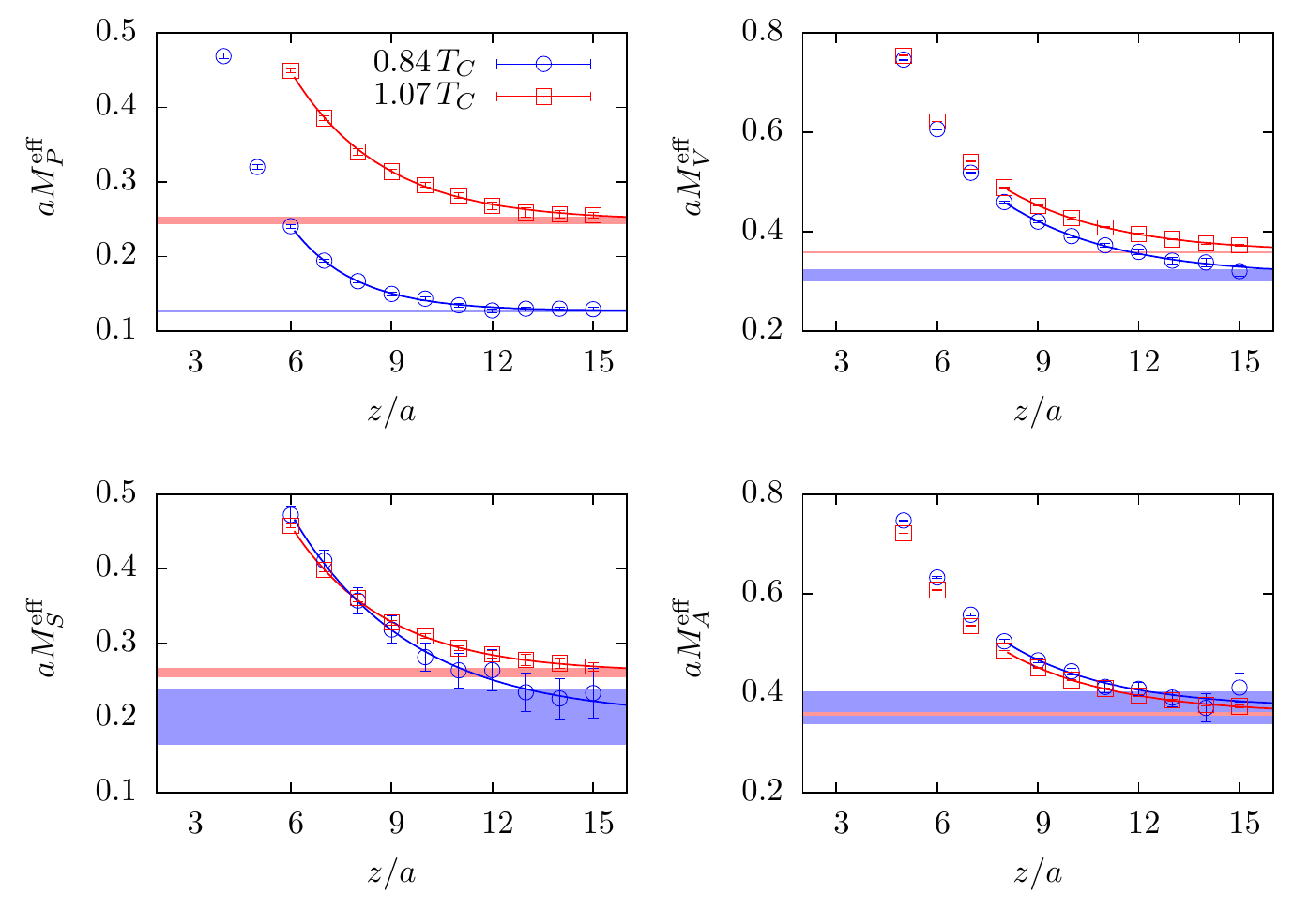}
 \caption{Effective masses for $T=0.84$ and $1.07\,\Tc$ in $P$ (top left), $S$
(bottom left), $V$ (top right) and $A$ (bottom right) channels in lattice units
from scan \scan{C1}. The solid lines are the results from a constant plus
exponential fit to the points that are within the area for which the curves are
shown and the shaded area indicates the result for the effective mass.}
 \label{fig:scm-extract}
\end{figure}

The screening correlators have been measured in the $x_3$ direction on the
stored configurations using unsmeared point sources. To make efficient use of
the generated configurations we have computed the correlation functions for 48
randomly chosen source positions (see also~\cite{Brandt:2015sxa}). Compared
to Ref.~\cite{Brandt:2013mba}, we have thus enlarged the statistics by another factor
of three. Screening masses are extracted from the effective mass~\footnote{Note,
that the procedure differs from the one used in~\cite{Brandt:2014qqa}, where we
have used a direct fit to the correlator. In general, the two sets of results
are consistent within errors. The current set of measurements supersedes the
ones from~\cite{Brandt:2014qqa}.}, $aM_X^{\rm eff}$, defined by the formula for
the inverse hyperbolic cosine,
\be
\label{eq:cosh-effmass}
aM_X^{\rm eff}(z) = \ln \left[ \frac{C(z+a) + C(z-a)}{2C(z)} +
\sqrt{ \left( \frac{ C(z+a)+C(z-a)}{2C(z)} \right)^2 -1 } \right]
\,,
\ee
where $C=C_{XX}$. Since in scans \scan{C1} and \scan{D1} the spatial extent is
rather small we could not find reasonable plateaus in most of the cases due to
contaminations from excited states (see also~\cite{Brandt:2014qqa}). To take the
contaminations into account we have fitted the results for the effective mass to
the form
\be
\label{eq:effmass-fit}
aM_X^{\rm eff}(z) = aM_X + aA \exp\big( - \Delta z \big) \,,
\ee
where $A$ and $\Delta$ are additional fit parameters. In
figure~\ref{fig:scm-extract} we show examples for the effective masses in the
different channels above and below $\Tc$, together with the results from the
associated fits.

\begin{figure}[t]
 \centering
\includegraphics[]{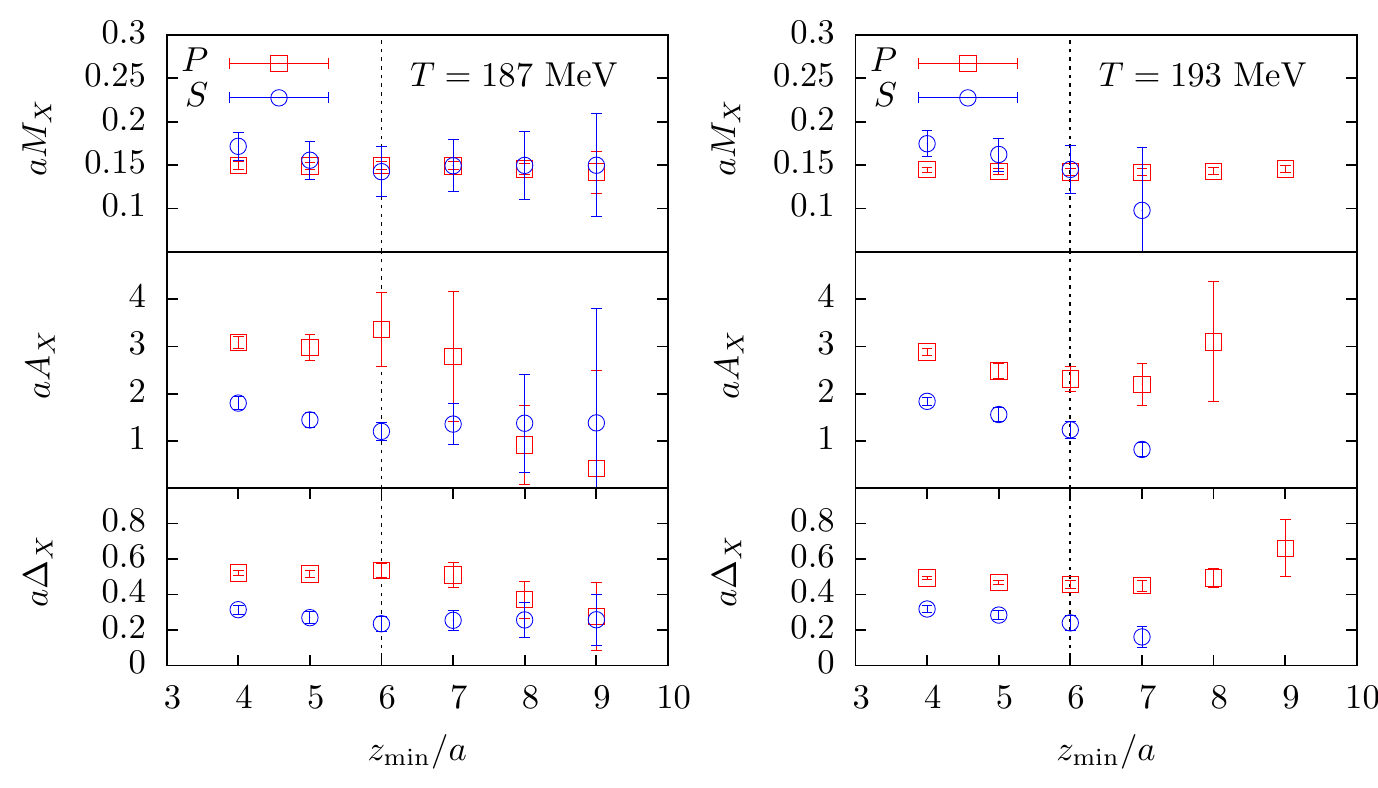}
 \caption{Result for the fit parameters of the fits for the extraction of
 screening masses versus the starting point $z_{\rm min}$ of the fit. Shown are
 the results for the temperatures of 187 MeV (left) and 193 MeV (right) from
 scan \scan{D1}. The vertical line corresponds to the choice for $z_{\rm min}$
 used to obtain the final results.}
 \label{fig:scm-frange}
\end{figure}

\begin{figure}[t]
 \centering
\includegraphics[]{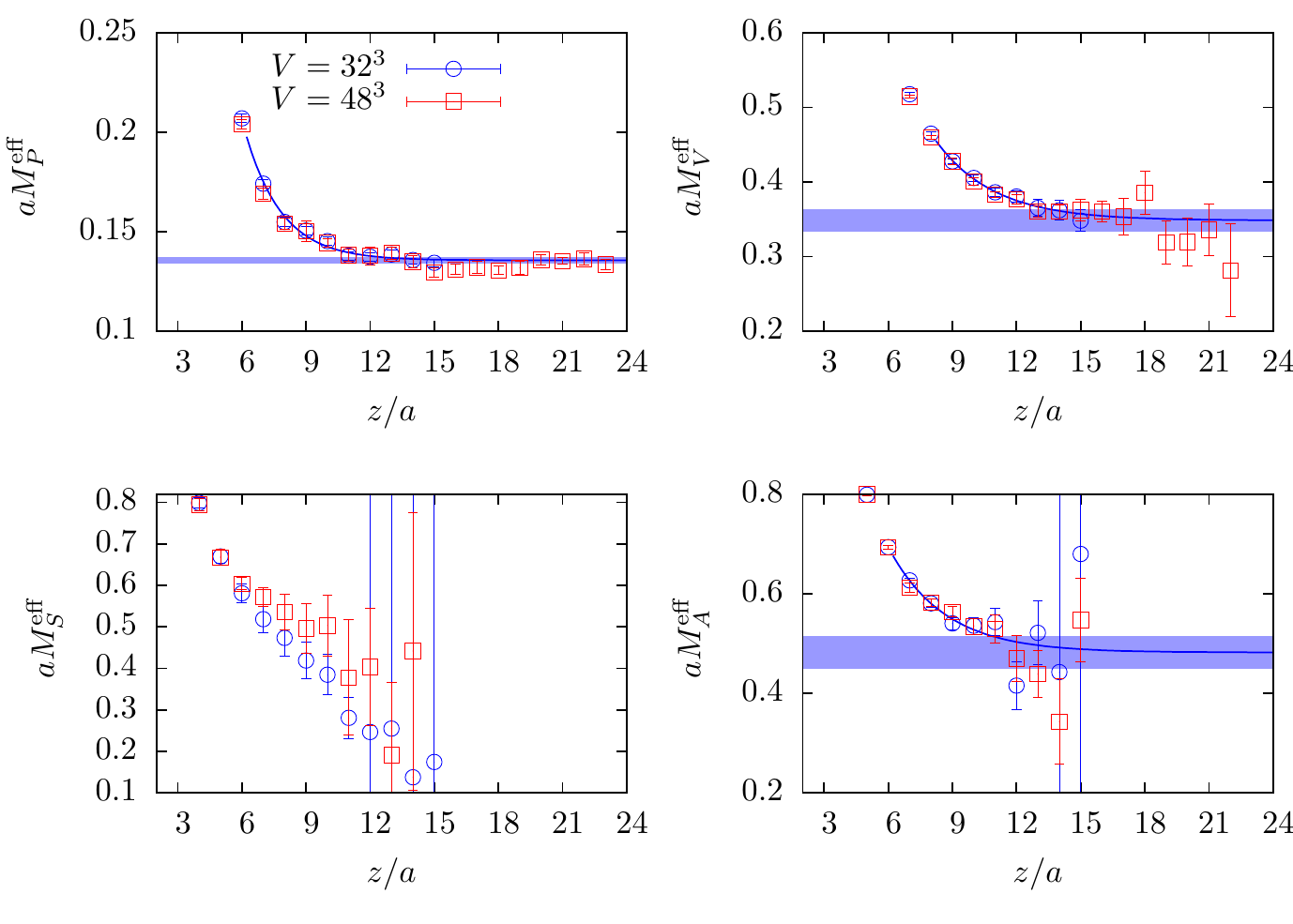}
 \caption{Effective masses for $T=150$ MeV for scan \scan{C1} with a $32^3$
volume and for the future scan \scan{C2} with a $48^3$ volume. The solid lines
are the results from a constant plus exponential fit to the points for scan
\scan{C1} that are within the area for which the curves are shown and the shaded
area indicates the result for the effective mass. For the scalar screening
masses the excited states fits did not work, so that we could not include any
fit result in the plot.}
 \label{fig:scm-finiteV}
\end{figure}

The formula for the effective mass follows to leading order when the
contamination from the first excited state is included in the correlation
function. In this case $\Delta$ represents the energy gap between groundstate
and first excited state in this channel. However, when contaminations from
higher excited states become important $\Delta$ (and $A$) will also contain
contributions from those. In both cases a fit to the form~\refc{eq:effmass-fit}
is known to improve the extraction of the groundstate energy significantly and
to remove most of the contaminations of the excited states. Note that neither
$A$ nor $\Delta$ are of direct importance for the following analysis. The fits
to the form~\refc{eq:effmass-fit} typically work well for a variety of fitranges
with starting points $z_{\rm min}$ in the range from $4a$ to $9a$ for $P$ and
$S$ channels and between $6a$ and $11a$ for $V$ and $A$ channels.~\footnote{For
fits with yet larger $z_{\rm min}$ the information in the data is not sufficient
to constrain all three fit parameters sufficiently. In this case one could try a
fit to a constant to extract $M_X$, but this procedure is less reliable than
the use of eq.~\refc{eq:effmass-fit}.} In these regions the results for $M_X$ do
not depend significantly on the particular choice of fitrange and also the
change of $A$ and $\Delta$ is only significant in a few cases. We show the
results for $M_X$, $A$ and $\Delta$ from two representative cases in $P$ and
$S$ channels close to the critical temperature for \scan{D1} (where
contaminations from excited states are typically expected to be most pronounced
due to the small quark mass in the scan) in figure~\ref{fig:scm-frange}. Note,
that the $S$ channel is the more problematic one, both due to large statistical
uncertainties and large contaminations from excited states (see also
figures~\ref{fig:scm-extract} and~\ref{fig:scm-finiteV}). In fact, the results
shown in the right panel of figure~\ref{fig:scm-frange} for the $S$ channels are
an example for the case where the fits do not work starting from $z_{\rm
min}/a=8$. Given these results, we conclude that for these values of $z_{\rm
min}$ the main contribution to $\Delta$ comes from the first excited state.
Within these regions where the fitparameters are insensitive to the particular
choice for $z_{\rm min}$ we can choose to work with any value of $z_{\rm
min}$. We decided to use $z_{\rm min}=6a$ for $P$ and $S$ channels and $z_{\rm
min}=8a$ for $V$ and $A$ channels. For these values all of the fits give a
reasonable $\chi^2/$dof around 1.~\footnote{Note that there were a few cases for
the $S$ and $A$ channels for $T<T_C$ where the fitranges needed to be adapted
due to large uncertainties for larger $z$ values. However, those cases have no
influence on the further analysis.}

\begin{figure}[t]
 \centering
\includegraphics[]{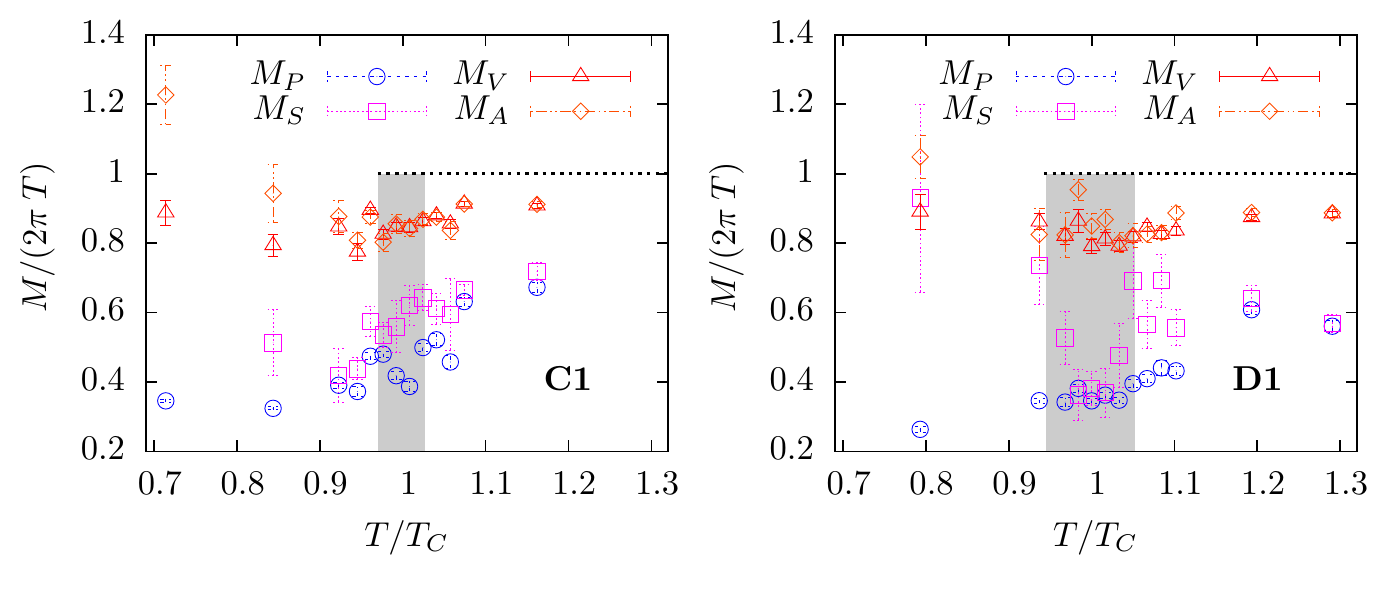}
 \caption{Results for the screening masses in the different channels for scans
\scan{C1} (left) and \scan{D1} (right). The screening masses are normalised to
the asymptotic limit $M^\infty$.}
 \label{fig:scm-results}
\end{figure}

The results for the screening masses are shown in figure~\ref{fig:scm-results}.
At $T/\Tc\approx0.7$ the screening masses show the expected splitting from the
zero temperature meson masses~\cite{Karsch:2003jg}. While the masses in the $P$
and $V$ channels initially remain constant, indicated by a slight decrease of
$M/T$, the screening masses in $S$ and $A$ channels decrease drastically in the
approach to $\Tc$. Around $T/\Tc\approx0.9$ all screening masses start to
increase. In particular, the screening masses in $P$ and $S$ channels are
drastically enhanced. Around $\Tc$ the screening masses in the $V$ and $A$
channels are mostly degenerate and around 85 to 90\% of the asymptotic $2\pi T$
limit, independent of the quark mass in the scan. This is consistent with the
findings in simulations with staggered 
fermions~\cite{Cheng:2010fe,Banerjee:2011yd}.

The screening masses in $P$ and $S$ channels move closer together and fluctuate
strongly around $\Tc$. For the $P$ channel the screening masses are only around
40 to 50\% of the $2\pi T$ limit for scan \scan{C1} with a pion mass of around
300~MeV and 35 to 40\% for scan \scan{D1} with a pion mass around 200~MeV,
indicating a quark mass dependence of the properties of pseudoscalar (and
scalar) states around $\Tc$. The screening mass in the $S$ channel is typically
around 10\% larger than $M_P$. In the temperature interval covered by our
calculations above $\Tc$, all screening masses are below the asymptotic
high temperature limit. Note that
weak-coupling calculations~\cite{Laine:2003bd,Alberico:2007yj}
predict the asymptotic approach to occur from above, implying
that the screening masses must cross the value $2\pi T$ at a certain
temperature.

\begin{figure}[t]
 \centering
\includegraphics[]{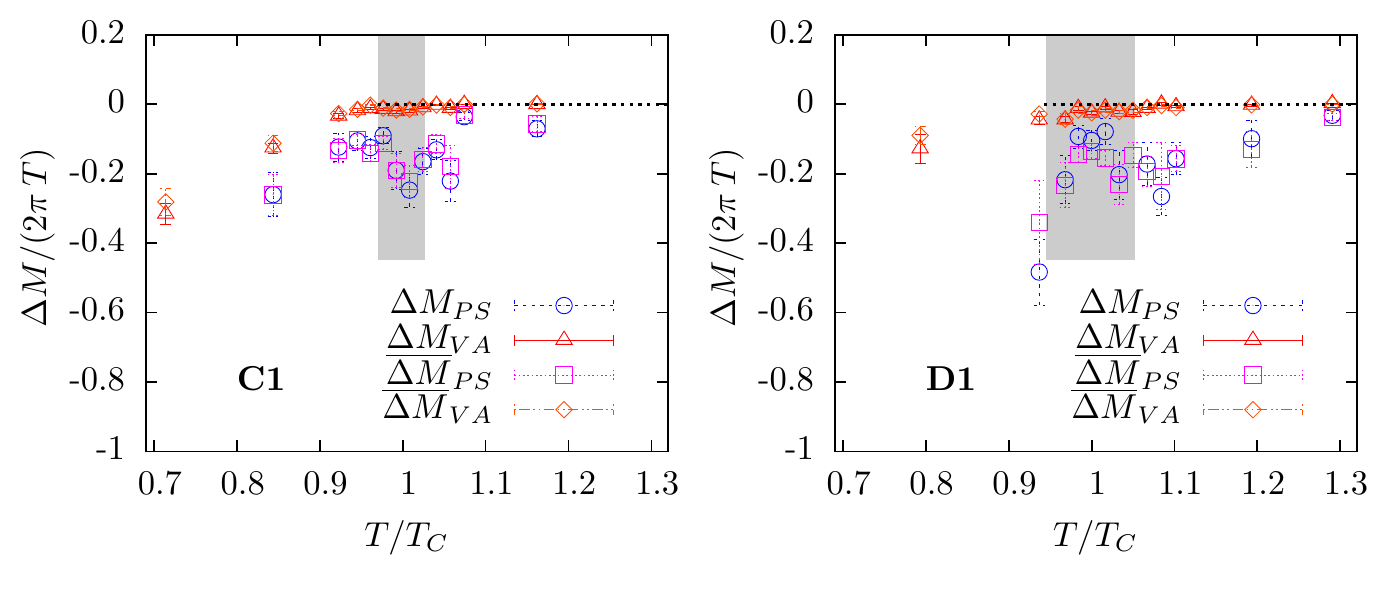}
 \caption{Results for the screening mass differences in the $P$ and $S$
channels, $\Delta M_{PS}$, and in $V$ and $A$ channels, $\Delta M_{VA}$, for
scans \scan{C1} (left) and \scan{D1} (right). The differences are
normalised to $2\pi T$.}
 \label{fig:scm-differences}
\end{figure}

Our findings are in good agreement with the results for screening
masses obtained with staggered
fermions~\cite{Cheng:2010fe,Banerjee:2011yd}. In simulations within the
quenched approximation, finite size effects have been found to be significant
up to aspect ratios of $N_s/N_t=4$~\cite{Wissel:2005pb,Muller:2013}. To get an idea
about their magnitude, we
compare the effective masses in the different channels at $T=150$~MeV from scan
\scan{C1} (with a volume of $32^3$) with those obtained from a simulation
with the same parameters but an increased spatial volume of $48^3$. The
comparison is shown in figure~\ref{fig:scm-finiteV}, 
no significant finite size effects are visible in the data. Note that
the comparison is done for the lattice with the smallest value of $M_P L$
(which is the relevant quantity governing the size of finite volume effects in the confined 
phase) in the scan \scan{C1}, being smaller than $M_P L$ for
all simulation points in the transition region. We thus conclude that our final
results are not strongly affected by finite size effects.

The chiral symmetry restoration pattern can be investigated by the degeneracies
of the screening masses. To extract the differences,
\be
\label{eq:diff-def}
\Delta M_{YX} = M_Y - M_X \,,
\ee
we have used the plateau in the effective masses of the ratios of the two
correlation functions. In these ratios some of the fluctuations between
different ensembles, evident in figure~\ref{fig:scm-results}, and statistical
fluctuations cancel. For these differences, some of the contaminations of the
excited states cancel as well, so that we could use a fit to a constant, where
we reduced the fitrange to the last few points. As for the fits to extract the
screening masses, we have explicitly checked that our results do not depend
significantly on the particular choice for the fitrange. This is in particular
true when we consider that the main uncertainties in the following analysis are
coming from the fluctuation between different ensembles in the region close to
$T_C$. On top of these checks for the fits, we also checked the extraction of
the mass splittings using a fit to the plateau in the ratio of effective masses
$M_X/M_Y$ to obtain another estimate for the difference via
\be
\label{eq:alt-diff-fit}
\overline{\Delta M}_{YX} \equiv M_Y \Big( \frac{M_X}{M_Y} - 1 \Big) .
\ee
In addition, we have also compared the results from the direct differences of
screening masses. Note that in a very few cases the fit for $\Delta M_{PS}$
failed, and in one case this also happened within the transition region
($\beta=5.37$ for scan \scan{D1}). We have excluded this datapoint from the
following analysis, but we have checked with the mass difference of the
independently determined screening masses that $\Delta M_{PS}$ for this $\beta$
value indeed lies within the uncertainty of the final estimate.

In figure~\ref{fig:scm-differences} we show the results for the mass
differences. The plot indicates the degeneracy of $M_V$ and $M_A$ and the
associated restoration of the $\SU_A(2)$ symmetry around $\Tc$ for both scans.
As already noted above, this is in agreement with the results from staggered
fermion simulations~\cite{Cheng:2010fe,Banerjee:2011yd}. This also adds to the
confidence concerning the extracted transition temperatures. In particular, both
estimates for the mass difference agree very well over the whole temperature
range for these channels (the ratio fit does not work for the lowest
temperatures so that we have no reliable results in that region). For $M_P$ and
$M_S$ there is still a significant mass splitting in the region around $\Tc$,
which, however, appears to become weaker with decreasing quark mass. For the
following analysis we will conventionally use the direct results for the mass
differences, i.e. the results labeled by $\Delta M_{XY}$. The persistent mass
splitting in $P$ and $S$ channels implies a residual breaking of $\Ua$ around
$\Tc$ in agreement with what has been found with staggered~\cite{Cheng:2010fe},
domain wall~\cite{Bazavov:2012qja,Bhattacharya:2014ara,Cossu:2014aua} and
overlap~\cite{Cossu:2013uua,Tomiya:2014mma} fermion formulations. To make any
statement about the fate and strength of the breaking in the chiral limit,
however, we still need to perform a chiral extrapolation, which is the topic of
the next section.

\subsection{On the relative size of the $\Ua$ breaking effects around $T_C$}
\label{sec:ua1-breaking}

An important question is which amount of symmetry breaking is ``strong'' or
``weak''. When looking at the
domain wall fermion results~\cite{Bazavov:2012qja,Bhattacharya:2014ara} for
chiral susceptibilities,
one might be led to the conclusion that the breaking is significant
in the chiral limit. The same is true if one looks at the eigenvalues of the
associated Dirac operator~\cite{Cossu:2014aua}. However, the chiral
susceptibilities include contributions from contact terms which might give an
additional contribution that overwhelms the effect of chiral symmetry breaking.
The eigenvalues are a more sensitive probe of the $\Ua$ breaking. In this case,
studies with the overlap operator~\cite{Cossu:2013uua,Tomiya:2014mma,Cossu:2015lnb} 
have indicated that the residual breaking in the chiral limit might be weak in
contrast to the breaking implied by the eigenvalues of the domain wall operator
at finite extent of the fifth dimension.

In contrast to the studies discussed above, we use non-perturbatively
O($a$)-improved Wilson fermions, which break chiral symmetry
explicitly.  The symmetry only becomes restored in the continuum
limit.  Since we use the mesonic screening spectrum as a probe, the
results are expected to approach the continuum with O($a^2$)
corrections. On our relatively fine lattices ($N_t=16$), we
expect these effects to be numerically small. We saw in the previous
section that we do observe -- to a good accuracy -- the expected
degeneracy between the vector and axial-vector screening above
$T_C$, signalling the restoration of the non-anomalous chiral symmetry.
Similarly, if the $\Ua$ symmetry becomes effectively restored, we expect to
obtain $\Ua$-breaking mass splittings of $O(a^2)$. In particular, since both the
anomalous and the non-anomalous chiral symmetry are broken explicitly by the
Wilson term, we expect the lattice artefacts to be of the same order of
magnitude. Owing to the results for the difference between vector and
axial-vector screening masses we expect this effect to contribute corrections
of order 10~MeV. Furthermore, any accidental cancellation between lattice
artefacts and a mass splitting in the continuum can only happen on this scale.
Determining the strength of the breaking with a relative precision better
than a few MeV requires taking the continuum limit.

\begin{table}[t]
\begin{center}
\small
\begin{tabular}{c|ccc|cc}
\hline
\hline
 & \scan{B1}$_\kappa$ & \scan{C1} & \scan{D1} & $m_{ud}=0$ linear & $m_{ud}=0$
sqrt \\
\hline
$\Delta M_{PS}$ [MeV] & -272(33)(479) & -172(24)(158) & -121(23)(127) & -82(282)
 & 5(590) \\
\hline
\hline
\end{tabular}
\end{center}
\caption{Results for $\Delta M_{PS}$ at $T_C$ for the different scans. The first
error is statistical, the second reflects the systematic error owing to the
fluctuations in the transition region. The value labeled with ``$m_{ud}=0$
linear'' is the result from the linear chiral extrapolation to all three points,
as explained in the text, and the value labeled with ``$m_{ud}=0$
sqrt'' assumes a quark mass dependence proportional to $\sqrt{m_{ud}}$. The
uncertainty on the result of the chiral extrapolations contains the statistical
as well as the systematical uncertainty.}
\label{tab:deltaps}
\end{table}

We will now look at the chiral extrapolation of the mass difference $\Delta
M_{PS}$ in physical units. First of all, using the value
$m_{a0}=980(20)$~MeV from the particle data group~\cite{Agashe:2014kda} we
obtain as an estimate for the zero temperature reference value at the physical
point
\be
\label{eq:zero-T-physical-measure}
\Delta M^{T=0}_{PS}=-845(20) \: \text{MeV} \,.
\ee
This value might serve as an estimate for the effect in the mass difference 
when the breaking of $\Ua$ is substantial. However, this estimate is valid for
physical, non-zero light quark mass, while we are interested in its value in the
chiral limit. The pion mass vanishes in the chiral limit, so
that $\Delta M_{PS}^{T=0,m_{ud}=0} = - m_{a0}^{m_{ud}=0}$. Next we need 
an estimate for $m_{a0}^{m_{ud}=0}$, for which we use the following ansatz:
we assume that the difference between chiral limit and physical point is solely due
to the change of the constituent quark masses $m_q^{\rm const}$ in the meson and
compare with another ``iso-vector'' scalar particle in the review of the
Particle Data Group (PDG)~\cite{Olive:2016xmw}, namely the $K_0^*$, where one of
the u/d quarks is replaced by a strange quark. This results in the ansatz
\be
\label{eq:defC}
m_{a_0}-m_{K_0^*} = C (m_{u,d}^{\rm const}-m_s^{\rm const})\;,
\ee
where $C$ is a proportionality constant, which can be determined from
eq.~\refc{eq:defC}. Using $C$ we can estimate
the mass of the scalar in the chiral limit following
\be
m_{a_0} - m_{a_0}^{m_{ud}=0} = 2 C m_{u,d} \,,
\ee
where the factor 2 comes from the fact that we need to send the masses of two
quarks to zero. Using the numbers from the PDG~\cite{Olive:2016xmw} we obtain
the final estimate $m_{a0}^{m_{ud}=0}=945(41)$ MeV. The error estimate follows
from the uncertainties associated with the masses of the $a0$ and the $K_0^*$
mesons. This is a rather crude estimate, but it is unlikely that it
underestimates the effect by an order of magnitude (even then
$m_{a0}^{m_{ud}=0}\approx 600$ MeV, which does not change the picture
dramatically). Our final estimate for the chiral limit is
\be
\label{eq:zero-T-measure}
\Delta M^{T=0,m_{ud}=0}_{PS}=-945(41) \: \text{MeV} \,.
\ee

The width of the transition region must be taken into account when
we extract an estimate for $\Delta M_{PS}$ from our simulations. We thus compute
the difference from a fit to a constant to the data points in the grey bands in
figure~\ref{fig:scm-differences}. The spread of the results in the region is
taken as a systematic uncertainty on top of the statistical uncertainty of
the average. The results from this procedure are listed in
table~\ref{tab:deltaps}. Here we have also included a result for scan
\scan{B1}$_\kappa$ to be able to perform a sensible chiral extrapolation.
Unfortunately, \scan{B1}$_\kappa$ is not at fixed quark mass and thus remains
longer in the vicinity of $\Tc$, since the latter increases with the quark mass.
This accounts for the rather large error bars for the associated $\Delta
M_{PS}$.

\begin{figure}[t]
 \centering
\includegraphics[]{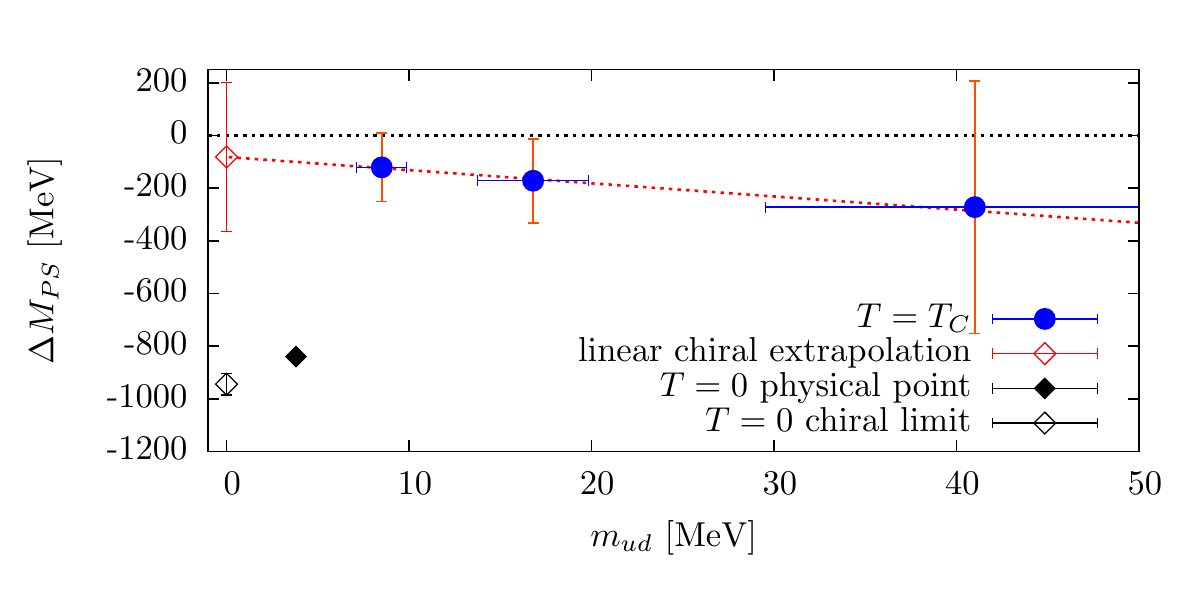}
 \caption{Chiral extrapolation of $\Delta M_{PS}$ in physical units in
comparison to the zero temperature pendant estimated as explained in the text.
The red line and the associated red point at $m_{ud}=0$ indicate the result
from a linear chiral extrapolation.}
 \label{fig:ua1-chext}
\end{figure}

To perform the chiral extrapolation for $\Delta M_{PS}$ we need to deduce
its quark mass dependence. Since the pion is a Goldstone boson, its mass is
expected to be proportional to $\sqrt{m_{ud}}$, at least at small temperatures,
$T<\Tc$. On the other hand, the mass of the scalar should depend linearly on the
quark mass which might also be the case for the pion at $\Tc$, where chiral
perturbation theory breaks down. Given that we have only three data points at
our disposal with relatively large uncertainties, our data clearly does not
allow for a detailed investigation of the quark mass dependence of $\Delta
M_{PS}$. We thus perform two types of fits; (i) linear in $m_{ud}$, (ii)
proportional to $\sqrt{m_{ud}}$. The results for the two different types of fits
including all three data points are listed in table~\ref{tab:deltaps}. We see
that both results are consistent with zero within the relatively large error
bars. As our final estimate we will thus use the linear fit. The associated
result is shown in figure~\ref{fig:ua1-chext}. We have also checked the
robustness of the result with a linear fit using only the data from scans
\scan{C1} and \scan{D1}. For this chiral extrapolation the central value remains
within the error bar of the result quoted in table~\ref{tab:deltaps}, but the
uncertainty of the result increases.

Figure~\ref{fig:ua1-chext} indicates that the $\Ua$-breaking
screening-mass difference is a fairly small effect and has a mild quark-mass
dependence only. In fact, at $m_{ud}=0$ the breaking effects
are strongly suppressed compared to the effect in the same quantity at zero
temperature and consistent with zero. Our result is thus in qualitative
agreement with the results from the spectrum of overlap
fermions~\cite{Tomiya:2014mma,Cossu:2015lnb}.

\section{Conclusions}
\label{sec:concl}

In this paper we studied the finite temperature transition of two-flavour QCD on
$16\times32^3$ (and $48^3$) lattices. In
particular, we have presented our results for the deconfinement and chiral
symmetry restoration temperatures, extracted from the inflection point of the
Polyakov loop and the peak in the susceptibility of the subtracted chiral
condensate, respectively. In agreement with previous studies in the literature, we find both
temperatures to decrease with the quark
mass~\cite{Karsch:1993tv,Karsch:1994hm,Aoki:1998wg,Bernard:1999xx,
DElia:2005bv,Cossu:2007mn,Bonati:2009yg,AliKhan:2000iz,Bornyakov:2009qh,
Bornyakov:2011yb,Burger:2011zc}. Our results for the chiral symmetry
restoration temperatures, reported in table~\ref{tab:trans-points}, are within uncertainties
consistent  with the values quoted by the tmfT
collaboration~\cite{Burger:2011zc,Burger:2014xga}.

The ultimate goal of our ongoing efforts is to determine the order of the transition in the chiral
limit. In an attempt to extract information about the chiral limit, we tested for scaling of
the transition temperatures in the approach to the chiral limit. As already
reported in~\cite{Brandt:2013mba}, the scaling behaviour alone is not
sufficiently constrained to distinguish between a second order O(4) chiral transition or
a first order transition with a Z(2) endpoint. This is consistent
with earlier findings of the tmfT collaboration~\cite{Burger:2011zc}, even
though we were able to reduce the pion masses in our study down to 200~MeV.
Extrapolations of the critical temperatures to the chiral limit are not
very sensitive to the universality class. Our results, when interpreted in terms
of the O(4) and U(2) scenarios are consistent with the findings from the tmfT
collaboration for O(4) scaling~\cite{Burger:2011zc}, and somewhat smaller than
those for Wilson fermions at $N_t=4$~\cite{AliKhan:2000iz} and the QCDSF-DIK
collaboration with different $N_t$ values~\cite{Bornyakov:2011yb}.
The fact that the chiral transition temperature does not appear to be very
sensitive to the universality class used for the chiral extrapolation is one
particular manifestation of the difficulty to
distinguish between different scenarios. Even a reduction of the
error bars by an order of magnitude in combination with results from smaller
quark masses might not be sufficient to allow to distinguish between the
universality classes when using the scaling of the transition
temperatures alone.

As an alternative, we have investigated the strength of the anomalous breaking of the $\Ua$
symmetry in the chiral limit by
computing the symmetry restoration pattern of screening 
masses in various isovector channels. At $T/\Tc\approx0.7$ the
screening masses assume values close to the zero temperature meson
masses. Initially the masses in the scalar and axial-vector channels
decrease, before at $T/\Tc\approx0.9$ all masses start to increase.
Around $\Tc$ the screening masses in the vector and axial-vector
channels are degenerate and about 85 to 90\% of the asymptotic $2\pi
T$ limit, while the screening masses in the pseudoscalar channel are
about 35 to 50\% of this limit and exhibit a significant dependence
on the quark mass. The screening mass in the scalar channel is
typically around 10\% larger. When going to higher temperatures all
screening masses approach the $2\pi T$ limit from below, 
while leading-order weak-coupling calculations~\cite{Laine:2003bd,Alberico:2007yj}
predict an asymptotic approach from above. All these findings are in
qualitative agreement with the results found in $N_f=2+1$ simulations
with staggered fermions~\cite{Cheng:2010fe,Banerjee:2011yd}.

To quantify the strength of the $\Ua$-anomaly
in the chiral limit, we use the difference between scalar and
pseudoscalar screening masses, $\Delta M_{PS}$. Unlike susceptibilities,
screening masses probe exclusively the long distance properties of the
correlation functions and thus do not suffer from contact terms. The comparison
of the chirally extrapolated value, $\Delta M^{m_{ud}=0}_{PS}=-81(282)$ MeV to
its zero temperature analogue $\Delta M^{T=0,m_{ud}=0}_{PS}=-945(41)$ MeV (cf.
eq.~\refc{eq:zero-T-measure}) suggests that the 
$\Ua$-breaking is strongly reduced at the transition temperature (see also
figure~\ref{fig:ua1-chext}). If this effect persists in the continuum limit, it
disfavours a chiral transition in the $O(4)$ universality class.

\section*{Acknowledgments}

We thank our colleagues from CLS for the access to the zero-temperature
ensembles. B.B. would like to thank Bastian Knippschild for sharing his routine
for APE smearing and Gergely Endr\H{o}di for many helpful discussions. We
acknowledge the use of computer time for the generation of the gauge
configurations on the JUROPA, JUGENE and JUQUEEN computers of the Gauss Centre
for Supercomputing at Forschungszentrum J\"ulich, allocated through the John von
Neumann Institute for Computing (NIC) within project HMZ21. The correlation
functions and part of the configurations were computed on the dedicated QCD
platforms ``Wilson'' at the Institute for Nuclear Physics, University of Mainz,
and ``Clover'' at the Helmholtz-Institut Mainz. We also acknowledge computer
time on the FUCHS cluster at the Centre for Scientific Computing of the
University of Frankfurt. This work was supported by the {\em Center for
Computational Sciences in Mainz} as part of the Rhineland-Palatine Research
Initiative and by DFG grant ME 3622/2-1 {\em Static and dynamic properties of
QCD at finite temperature}. B.B. has also received funding by the DFG via
SFB/TRR 55 and the Emmy Noether Programme EN 1064/2-1.

\appendix

\section{Simulation and analysis details}

\subsection{Simulation algorithms and associated constraints}
\label{app:sim-algos}

The simulations have been done using the deflation accelerated versions of the
Schwarz~\cite{Luscher:2005rx,Luscher:2007es} (DD) and
mass~\cite{Hasenbusch:2001ne,Marinkovic:2010eg} (MP) preconditioned HMC
algorithms. Both algorithms employ the Schwarz preconditioned and deflation
accelerated generalised conjugate residual (DFL-SAP-GCR) solver introduced by
L\"uscher in~\cite{Luscher:2003qa,Luscher:2007se}. Due to the efficient solver,
both algorithms exhibit an improved scaling behaviour when lowering the
quark mass and the lattice spacing.

The block structures used in the solver and the preconditioning of the HMC
algorithm impose constraints on the size of the local lattices allocated to the single processors.
The main limitation follows from the constraint that Schwarz preconditioning
blocks in the solver (SAP-blocks) have a minimal size of
$4^4$~\cite{Luscher:2003qa}. At least two of these blocks need to fit into a
sublattice, so that the minimal sublattice size when using a SAP solver is
$8\times4^3$. This restriction translates directly to the version of the MP-HMC
algorithm from~\cite{Marinkovic:2010eg}. The restrictions for the sizes of the
DD-blocks are essentially the same as for the SAP-blocks, but it also affects
the efficiency of the HMC preconditioning, since the separation of modes works
most efficiently if the DD-blocks have a minimal size of about half a fermi in
each direction~\cite{Luscher:2005rx}. The inverse critical temperature
corresponds to a temporal extent of about 1~fm, meaning that the size of
the DD-blocks should be half the size of the temporal extent. For our scans the
optimal sublattice size for the DD-HMC is thus given by $16\times8^3$. Further
constraints follow from the use of even/odd preconditioning in the different
levels of the solver (see~\cite{Luscher:2003qa,Luscher:2007se} for the details).
Summarizing, the algorithm restricts one to use lattice sizes of
$N_{t/s}=8,12,16,20,\ldots$.

In our initial runs (see~\cite{Brandt:2010uw,Brandt:2010bn,Brandt:2012sk}) we
 used the DD-HMC algorithm. A drawback of the algorithm is that a
sizeable fraction $R$ of the links remain fixed during the molecular dynamics
evolution~\cite{Luscher:2005rx}. While the ergodicity can be restored by
shifting the gauge field between two HMC trajectories, the autocorrelation
between two trajectories increases significantly. For typical blocksizes of
$4^4$ and $8^4$ one obtains $R=0.09$ and 0.37. For $\Nt=16$ the optimal block
size is $8^4$, for which the autocorrelations are expected to be enhanced by
about a factor of 3 compared to the MP-HMC algorithm. In particular on large
scale machines, one is often forced to use a large number of processors and
decrease the local system size, meaning that autocorrelations increase
drastically and the algorithm becomes inefficient. An example is the
\scan{B3}$_\kappa$ lattice discussed in~\cite{Brandt:2013mba}.

Due to the suppression of low modes above $\Tc$ accompanying the effective
restoration of the chiral symmetry, the main motivation for using deflation
gradually disappears. The reduced benefit of deflation can be
observed in practice. However, for the lattices and quark masses in this study,
there is still a significant acceleration even above $\Tc$. In the region around
and below $\Tc$ we had rare appearances of problems with deflation, most likely
stemming from the fact that the ``little'' Dirac operator
(see~\cite{Luscher:2007se} for the details) has large condition numbers. Mostly
these issues could be solved by changing the parameters of the deflation
subspace.

\subsection{Error analysis}
\label{app:error-ana}

For the error analysis we employ the bootstrap procedure~\cite{Efron:1979} with
1000 bins. The data for different time slices of correlation functions is
correlated, which means that in fits the full covariance matrix has to be taken
into account. In practice, however, the uncertainties for the entries of the
correlation matrix are often not precise enough for a stable least-square
minimisation (see~\cite{Michael:1993yj} for instance). This is in particular
true for screening masses that typically have a bad signal-to-noise ratio. All
our fits to correlation functions have thus neglected the non-diagonal terms of
the correlation matrices.

For scale setting and renormalisation we have used interpolations of quantities
that have their own uncertainties and for which we have no access to the raw
data. To take these uncertainties into account, we have generated uncorrelated
pseudo-bootstrap distributions with 1000 bins, whose width reproduces the quoted
uncertainties. The interpolations have then been done for each bin, giving bins
for the desired quantities at each value of the coupling.

\subsection{Interpolation of zero temperature quantities}
\label{app:T0-interpol}

\begin{table}[t]
  \centering
  \begin{tabular}{c|ccc}
    \hline
    \hline
    $\beta$ & 5.20 & 5.30 & 5.50 \\
    \hline
    \hline
    $r_0/a$ & 6.15( 6) & 7.26( 7) & 10.00(11) \\
    $\kappa_c$ & 0.136055(4) & 0.136457(4) & 0.1367749(8) \\
    \hline
    \hline
  \end{tabular}
  \caption{Input from zero temperature for scale setting and the estimation of
           LCPs taken from~\cite{Fritzsch:2012wq}.}
  \label{tab:T0_input}
\end{table}

\begin{figure}[t]
 \centering
\includegraphics[]{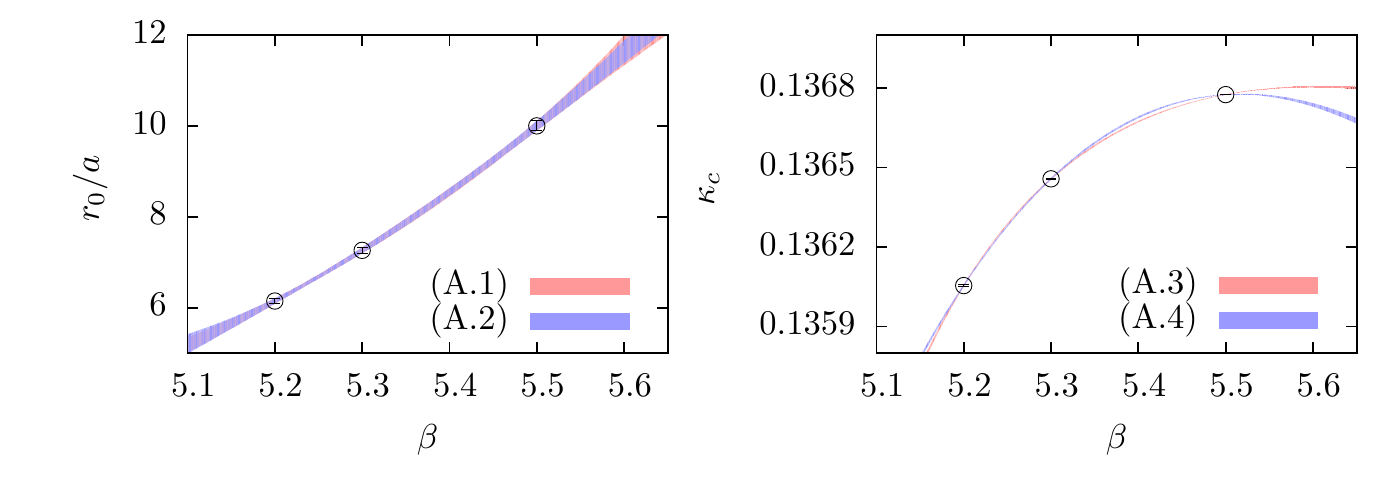}
 \caption{Interpolations of $r_0/a$ (left) and $\kappa_c$ (right).}
 \label{fig:interpols}
\end{figure}

In this appendix we discuss the interpolations of several quantities, such as
lattice spacing and quark masses. For scale setting we perform an interpolation
of $r_0/a$ obtained from the CLS lattices~\cite{Fritzsch:2012wq,Leder:2010kz},
summarised in table~\ref{tab:T0_input}. The final result is obtained using the
ansatz~\cite{Guagnelli:1998ud}
\be
\label{eq:r0-interpol}
\ln[a/r_0(\beta)] = c_0 + c_1 \beta + c_2 \beta^2 \,,
\ee
motivated by the solution of the renormalisation group equation for the bare
coupling to two-loops. To check the model dependence of the result we use a
simple polynomial of second order,
\be
\label{eq:r0-interpol-check}
r_0/a(\beta) = \bar{c}_0 + \bar{c}_1 \beta + \bar{c}_2 \beta^2 \,.
\ee
The results for the coefficients are tabulated in
table~\ref{tab:mud-interpol-paras} and the interpolation is visualised in
figure~\ref{fig:interpols} (left). The plot displays the good agreement between
the two interpolations, indicating that in the region $5.20\leq\beta\leq5.50$,
systematic errors due to the ansatz for $r_0/a(\beta)$ are negligible.

Another important quantity is the critical hopping parameter $\kappa_c$. We
interpolate the result from~\cite{Fritzsch:2012wq}, listed in
table~\ref{tab:T0_input}, using the two different ans\"atze,
\begin{eqnarray}
\label{eq:kc-interpols}
 \kappa_c(\beta) & = & \frac{1}{8} \,
\frac{1+d_0/\beta+d_1/\beta^2}{1+d_2/\beta} \,, \\
 \kappa_c(\beta) & = & \frac{1}{8} +
\bar{d}_0/\beta+\bar{d}_1/\beta^2+\bar{d}_2/\beta^3 \,.
\end{eqnarray}
The results are also listed in table~\ref{tab:mud-interpol-paras} and the
interpolations are shown in figure~\ref{fig:interpols} (right). There is a
slight model dependence in the region $5.35\lesssim \beta \lesssim5.45$.
However, the interpolation mostly concerns the estimation of LCPs and does not
enter the final analysis.

For the other renormalisation factors and improvement coefficients we have used
known interpolation formulas from the literature: for $Z_V$ we have used the
interpolation formula from~\cite{DellaMorte:2005rd}; for $c_A$ the one from
\cite{DellaMorte:2005se}; for $Z_A$ and $Z_P$ we have used the results
from~\cite{Fritzsch:2012wq}; for $b_m$ and $[b_A-b_P]$ we have used the
non-perturbative result from~\cite{Fritzsch:2010aw}.

\subsection{Estimating lines of constant physics}
\label{app:lcps}

\begin{table}[t]
\begin{center}
\small
\begin{tabular}{cccc|cccc}
\hline
\hline
$c_0$ & $c_1$ & $c_2$ & & \hspace*{10mm} & $\bar{c}_0$ & $\bar{c}_1$ &
$\bar{c}_2$
\\
\hline
-13(17) & 3.9(62) & -0.21(58) &  & & 184(120) & -79(45) & 8.6(42) \\
\hline
\hline
$d_0$ & $d_1$ & $d_2$ & & & $\bar{d}_0$ & $\bar{d}_1$ & $\bar{d}_2$ \\
\hline
-4.089(19) & -2.967( 8) & -4.703(16) &  & & -0.82(4) & 10.2(4) & -29.0(9) \\
\hline
\hline
$z_0$ & $z_1$ & $z_2$ & \multicolumn{1}{c}{$\chi^2/$dof} & & & \\
\hline
55(37) & -20(14) & 1.9(13) & \multicolumn{1}{c}{0.2} & & & \\
\hline
\hline
\end{tabular}
\end{center}
\caption{Results of the fits for the interpolations of zero temperature results
and the analytic relation $am_{\rm PCAC}(\beta,\kappa)$. Note, that the for the
interpolations there are as many parameters as data points, so that it is a
parameterisation rather than a fit.}
\label{tab:mud-interpol-paras}
\end{table}

LCPs can be realised once we have an analytic relation between the bare
parameter $\kappa$ and the renormalised quark mass $m_{ud}$. This relation can
be obtained by using the two different definitions for $m_{ud}$ in the case of
Wilson fermions. The first definition uses the bare quark mass
$\bar{m}$~\cite{Luscher:1996sc},
\be
\label{eq:ren-mud-bare}
m_{ud}^{\rm bare} = Z_m \, ( 1 + b_m \, a\bar{m} ) \, \bar{m} \quad
\text{with} \quad a\bar{m} = \frac{1}{2} \, \Big( \frac{1}{\kappa} -
\frac{1}{\kappa_c} \Big) \,,
\ee
where $Z_m$ is the scheme dependent mass renormalisation factor and $b_m$ is
an $O(a)$ improvement coefficient. Another possibility is to
use the PCAC quark mass~\cite{Luscher:1996ug},
\be
\label{eq:ren-mud-pcac}
m_{ud}^{\rm PCAC} = \frac{Z_A}{Z_P} ( 1 + [ b_A - b_P ] \, a\bar{m} ) \, m_{\rm
PCAC} \,.
\ee
Here $Z_A$ and $Z_P$ are the renormalisation factors for the axial current and
the pseudoscalar density ($Z_P$ is scheme dependent) and $b_A$ and $b_P$ are due
to improvement. $m_{\rm PCAC}$ is the bare quark mass defined by the PCAC
relation (see~\cite{Fritzsch:2012wq,Brandt:2013dua}).

Both quantities in eqs.~\refc{eq:ren-mud-bare} and~\refc{eq:ren-mud-pcac} are
estimates for the renormalized quark mass $m_{ud}$ and differ only in lattice
artifacts. From eqs.~\refc{eq:ren-mud-bare} and~\refc{eq:ren-mud-pcac} we can
thus infer (see also~\cite{Fritzsch:2010aw})
\be
\label{eq:rel-bare-pcac}
am_{\rm PCAC} = Z_{\rm PCAC}(\beta) \, ( 1 + [b_m+b_P-b_A] \, a\bar{m} ) \,
a\bar{m} \,.
\ee
Eq.~\refc{eq:rel-bare-pcac} is valid up to $O(a^2)$ corrections and $Z_{\rm
PCAC}(\beta)$ only depends on the regularisation scheme (i.e. the lattice
discretisation), but not on the renormalisation scheme. Inserting the relation
between $a\bar{m}$ and $\kappa$ we obtain
\be
\label{eq:mpcac-vs-kappa}
am_{\rm PCAC}(\beta,\kappa) = \frac{Z_{\rm PCAC}(\beta)}{2} \, \Big(
\frac{1}{\kappa} - \frac{1}{\kappa_c(\beta)} \Big) + Z_2(\beta) \, \Big(
\frac{1}{\kappa} - \frac{1}{\kappa_c(\beta)} \Big)^2 + \,\ldots \,.
\ee
For $\kappa_c(\beta)$ we take the known relation from
appendix~\ref{app:T0-interpol}. When $m_{ud}$ is small
$a\bar{m}$ is small as well. Since the factor $Z_2$ contains the combination of
$b$-factors from eq.~\refc{eq:rel-bare-pcac}, which is dominated by $b_m$, and
is a number smaller than one, we can expect that the second term in 
eq.~\refc{eq:mpcac-vs-kappa} is negligible for small quark masses. We thus have
to obtain only the $\beta$-dependence of the factor $Z_{\rm PCAC}$. To this end
we use the data for $m_{\rm PCAC}$ at $T=0$
from~\cite{Fritzsch:2012wq,Brandt:2013dua}, listed in table~\ref{tab:T0_cond}.
Since data is available only for three couplings, we will use a simple second
order polynomial,
\be
\label{eq:zpcac-func}
Z_{\rm PCAC}(\beta) = z_0 + z_1 \beta + z_2 \beta^2 \,.
\ee
The fit with this ansatz works reasonably well, giving a small $\chi^2/{\rm
dof}$ of 0.2. In fact, $\chi^2/{\rm dof}$ does not make any statements about the
goodness of the ansatz for $Z_{\rm PCAC}(\beta)$, but it shows that for the
given range of quark masses $am_{\rm PCAC}$ and $a\bar{m}$ are indeed linearly
related. The results for the coefficients $z_i$ are listed in
table~\ref{tab:mud-interpol-paras}. For renormalisation we can then use
eq.~\refc{eq:ren-mud-pcac} with the constants from
appendix~\ref{app:T0-interpol} which are valid for the Schr\"odinger functional
scheme. To convert to the $\overline{\rm MS}$ scheme we use the conversion
factor $m^{\overline{\rm MS}}_{ud}/m^{\rm
SF}_{ud}=0.968(20)$~\cite{Brandt:2013dua,Brandt:2014qqa}.

\begin{figure}[t]
 \centering
\includegraphics[]{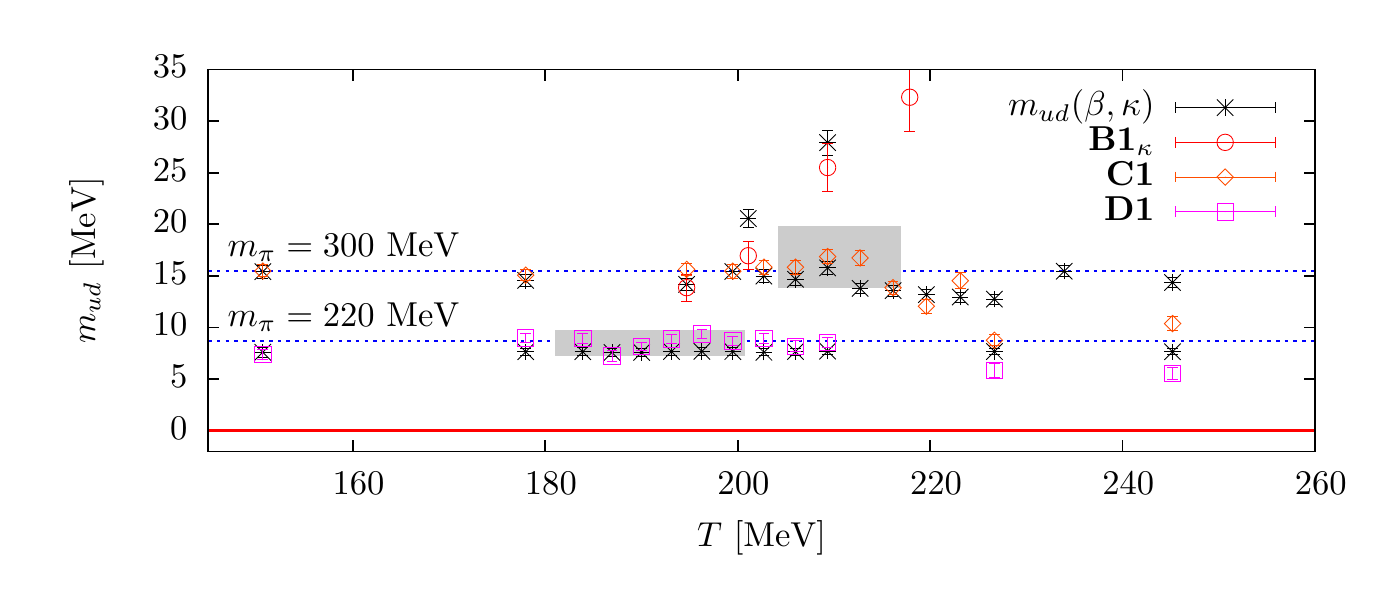}
 \caption{Comparison between $m_{ud}(\beta,\kappa)$ and the simulation results.}
 \label{fig:mud-checks}
\end{figure}

Note, that the final error bars on $m_{ud}$ obtained from the relation
$m_{ud}(\beta,\kappa)$ are much smaller than the uncertainties of the
coefficients suggest, due to compensations of changes in one parameter by the
others. Figure~\ref{fig:mud-checks} shows a comparison between the predictions
from $m_{ud}(\beta,\kappa)$ with the results obtained from the actual
simulations. The plot displays the good agreement for temperatures below and up
to $\Tc$. Note, that the tuning for scan \scan{C1} has been done using an
early version of the matching with less information from the $T=0$ side. This
explains the fact that the black points do not lie on a constant $m_{ud}$ line
in that case. The good agreement between the measured quark masses and the
predictions from the matching reported here indicates that this updated form of
the matching works well. Above $\Tc$, $m_{ud}$ is typically lower than expected.
It is unclear to us whether this result is an effect due to markedly different
cutoff effects above $\Tc$, or if the interpolation becomes worse at the higher
$\beta$ values. Note that our zero temperature ensembles are located at the
$\beta$-values corresponding to 150, 180 and 245 MeV, so that the systematic
error associated with the interpolation is expected to be most severe around 210
MeV.

\subsection{Interpolation of the zero temperature chiral condensate}
\label{app:chicond-inter}

\begin{table}[t]
\begin{center}
\small
\begin{tabular}{cc|cc|ccc}
\hline
\hline
$\beta$ & $\kappa$ & size & \# config & $am_{\rm PCAC}$ & $a^3\cond\bare$ &
$a^3\left<\overline{PP}\right>$ \\
\hline
5.20 & 0.13565 & $64\times32^3$ & 289 & 0.0158( 2) & 0.2727633(18) & 0.2064(
1) \\
 & 0.13580 & & 265 & 0.0098 ( 3) & 0.2727369(19) & 0.2203( 2) \\
 & 0.13590 & & 400 & 0.0057 ( 4) & 0.2727055(15) & 0.2428( 5) \\
 & 0.13594 & & 211 & 0.0044 ( 2) & 0.2726914(23) & 0.2609(10) \\
 & 0.13597 & $96\times48^3$ & 400 &  & 0.2726827( 7) & 0.2878( 9) \\
\hline
5.30 & 0.13610 & $64\times32^3$ & 162 &  & 0.2711093(16) & 0.1883( 1) \\
 & 0.13625 & & 508 & 0.0072 ( 3) & 0.2711386(11) & 0.2004( 2) \\
 & 0.13635 & &  & 0.00375(11) &  &  \\
 & 0.13638 & $96\times48^3$ & 250 & 0.00267( 9) & 0.2711641( 9) & 0.2438( 7)
\\
 & 0.13642 & $128\times64^3$ & 205 &  & 0.2711734( 5) &
0.3162(22) \\
\hline
5.50 & 0.13650 & $96\times48^3$ & 83 & 0.0091 ( 2) & 0.2685038( 8) & 0.1645(
1) \\
 & 0.13660 & & 188 & 0.0059 ( 2) & 0.2685683( 6) & 0.1702( 1) \\
 & 0.13667 & & 221 & 0.00343(12) & 0.2686063( 5) & 0.1782( 2) \\
 & 0.13671 & $128\times64^3$ & 137 & 0.00213( 6) & 0.2686293( 5) & 0.1912( 3) \\
\hline
\hline
\end{tabular}
\end{center}
\caption{Set of CLS $T=0$ lattices used for the determination of the relation
$m_{ud}(\beta,\kappa)$ and the computation of the condensate at zero temperature
and the associated results. For the measurements of $am_{\rm PCAC}$ and more
details on the lattices we refer to~\cite{Fritzsch:2012wq,Brandt:2013dua}.}
\label{tab:T0_cond}
\end{table}

For the computation of the renormalised condensate the computation of the $T=0$
subtraction terms is mandatory. Here we have computed $\cond\bare$ and
$\left<\overline{PP}\right>$ on the CLS lattices, summarised with the
measurement details in table~\ref{tab:T0_cond}. For the computation we have
used 12 $Z_2\times Z_2$ volume (for $\cond\bare$) and wall (for
$\left<\overline{PP}\right>$) sources on each configuration.

\begin{figure}[t]
 \centering
\includegraphics[]{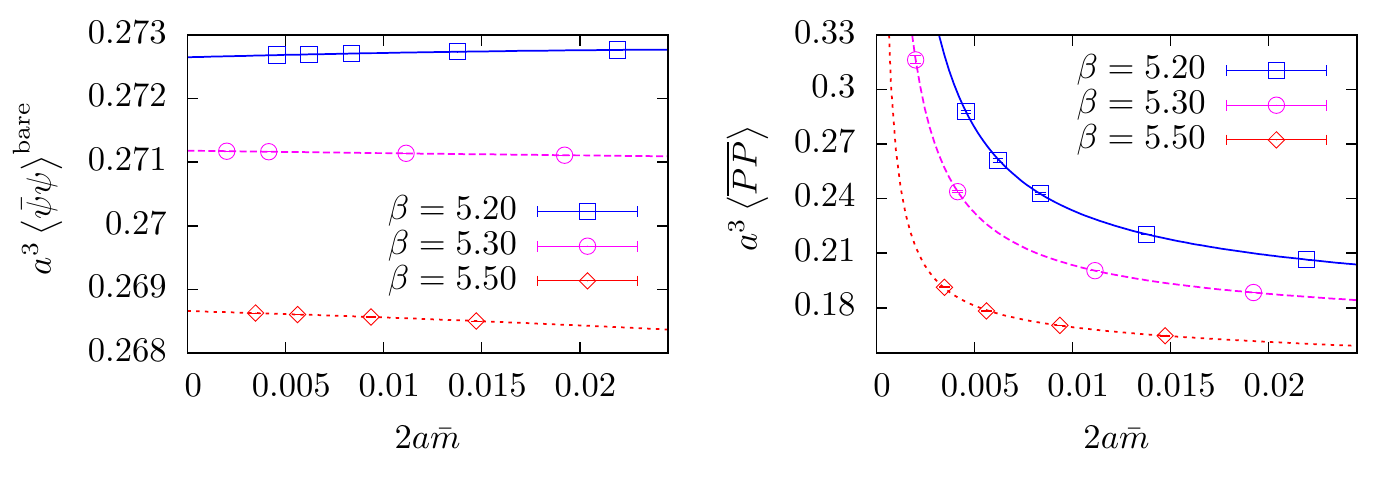}
 \caption{Interpolation of the results for the bare (left) and subtracted
(right) condensates obtained on the CLS lattices, as described in the text.}
 \label{fig:cond_interpol}
\end{figure}

We have tested several different functional forms for the interpolation and
found simple polynomials in $a\bar{m}$ to work best. Note, that
$\left<\overline{PP}\right>$ is expected to diverge linearly in
$a\bar{m}$, since the condensate, which is finite at $a\bar{m}=0$, is
proportional to $\bar{m}\left<\overline{PP}\right>$, which can be described by
a polynomial. We have thus used
\be
\label{eq:cond_interpol}
\begin{array}{rcl}
 \displaystyle a^3\cond\bare(\beta,\kappa) & = & \displaystyle p_1(\beta) +
p_2(\beta) (2a\bar{m}) + p_3(\beta) (2a\bar{m})^2 \vspace*{2mm} \\
 \displaystyle a^3\left<\overline{PP}\right>(\beta,\kappa) & = & \displaystyle
\bar{p}_1(\beta)/(2a\bar{m}) + \bar{p}_2(\beta) + \bar{p}_3(\beta) (2a\bar{m})
\,,
\end{array}
\ee
with
\be
\label{eq:pfacs}
p_i(\beta)=q^i_1 + q^i_2 \beta + q^i_3 \beta^2
\ee
(similarly for $\bar{p}_i$ with $q\to\bar{q}$) and as usual
$2a\bar{m}=1/\kappa-1/\kappa_c(\beta)$ with $\kappa_c(\beta)$ obtained from the
standard interpolation \refc{eq:kc-interpols}. Altogether, each fit has 9 fit
parameters. The results
are given in table~\ref{tab:cond_interpol_paras} and the interpolations for the
different $\beta$-values are shown in figure~\ref{fig:cond_interpol}. The
$\chi^2/$dof values are 3.3 and 5.5 for $\cond\bare$ and
$\left<\overline{PP}\right>$, respectively. The tiny error bars on
$\cond\bare$ and $\left<\overline{PP}\right>$ explain the
bad values for $\chi^2/$dof, despite the fit providing a satisfactory description of the data. Once
more, the uncertainties of the fit parameters do not reflect the uncertainty on
the results for the condensate. Note that we have neglected finite volume
effects for the condensates, which appears to be justified for $m_\pi
L\gtrsim4$.

\begin{table}[t]
\begin{center}
\small
\begin{tabular}{ccc|ccc|ccc}
\hline
\hline
$q^1_1$ & $q^1_2$ & $q^1_3$ & $q^2_1$ & $q^2_2$ & $q^2_3$ & $q^3_1$ & $q^3_2$ &
$q^3_3$ \\
\hline
0.539(4) & -0.087(2) & 0.0069(2) & 9.6(8) & -3.5(3) & 0.32(3) & -207(34) &
77(13) & -7.2(12) \\
\hline
\hline
$\bar{q}^1_1$ & $\bar{q}^1_2$ & $\bar{q}^1_3$ & $\bar{q}^2_1$ & $\bar{q}^2_2$ &
$\bar{q}^2_3$ & $\bar{q}^3_1$ & $\bar{q}^3_2$ & $\bar{q}^3_3$ \\
\hline
0.084(49) & -0.030(19) & 0.0027(17) & 6.1(55) & -2.1(21) & 0.19(19) & -80(170)
& 30(64) & -2.8(60) \\
\hline
\hline
\end{tabular}
\end{center}
\caption{Results for the fit parameters for the interpolation of the chiral
condensate at $T=0$.}
\label{tab:cond_interpol_paras}
\end{table}

\section{Simulation parameters and results}
\label{app:sim-paras}

\begin{table}[t]
\begin{center}
\small
\begin{tabular}{c|lrr|rrr}
\hline
\hline
Scan & $\beta$ & $\kappa$ & MDU & $T$ [MeV] & $m_{\rm ud}$ [MeV] & $\tau_{U_p}$
[MDU] \\
\hline
\hline
\vspace*{-2mm} & & & & & & \\
\scan{B1}$_{\rm \kappa}$ & 5.375 & 0.136500 & 10000 & 201(5) & 17.0(1.4) & 26(5)
\\
 & 5.400    &  & 22000 & 209(5) & 25.5(2.4) & 56(8) \\
 & 5.425    &  & 20200 & 218(5) & 32.3(3.4) & 32(4) \\
 & 5.450    &  & 20600 & 227(5) & 39.6(2.0) & 20(2) \\
 & 5.475    &  & 21600 & 236(5) & 39.3(2.0) & 22(3) \\
 & 5.48125  &  & 20200 & 238(5) & 46.2(2.2) & 18(2) \\
 & 5.4875   &  & 19600 & 240(6) & 46.9(2.1) & 21(3) \\
 & 5.49375  &  & 21400 & 243(6) & 45.3(2.1) & 19(3) \\
 & 5.496875 &  & 16200 & 244(6) & 47.9(2.2) & 16(2) \\
 & 5.500    &  & 21600 & 245(6) & 45.5(2.0) & 19(2) \\
 & 5.503125 &  & 16600 & 246(6) & 43.6(2.1) & 16(2) \\
 & 5.50625  &  & 21000 & 248(6) & 49.7(2.5) & 17(2) \\
 & 5.5125   &  & 21600 & 250(6) & 47.7(2.4) & 17(2) \\
 & 5.51875  &  & 19800 & 252(6) & 50.6(2.2) & 14(2) \\
 & 5.525    &  & 19400 & 255(6) & 53.5(2.5) & 17(2) \\
 & 5.550    &  & 19200 & 265(7) & 51.3(2.4) & 14(2) \\
 & 5.575    &  & 12200 & 275(8) & 55.6(2.7) & 13(2) \\
\hline
\scan{C1} & 5.20 & 0.135940 & 12320 & 151(4) & 15.5(6) & 96(28) \\
 & 5.30  & 0.136356 & 12260 & 178(4) & 15.1(6) & 24(3) \\
 & 5.355 & 0.136500 & 13040 & 195(5) & 15.6(7) & 34(7) \\
 & 5.37  & 0.136523 & 12560 & 199(5) & 15.5(7) & 19(3) \\
 & 5.38  & 0.136545 & 12240 & 203(5) & 15.8(7) & 25(4) \\
 & 5.39  & 0.136565 & 12240 & 206(5) & 15.9(7) & 20(3) \\
 & 5.40  & 0.136575 & 12080 & 209(5) & 16.9(7) & 24(5) \\
 & 5.41  & 0.136603 & 12320 & 213(5) & 16.8(7) & 31(5) \\
 & 5.42  & 0.136619 & 12080 & 216(5) & 13.9(7) & 23(4) \\
 & 5.43  & 0.136635 & 12480 & 220(5) & 12.1(7) & 22(4) \\
 & 5.44  & 0.136649 & 12166 & 223(5) & 14.5(8) & 15(2) \\
 & 5.45  & 0.136662 & 12000 & 227(5) &  8.8(6) & 15(2) \\
 & 5.50  & 0.136700 & 12480 & 245(6) & 10.4(8) & 14(2) \\
\hline
\scan{D1} & 5.20 & 0.135998 & 4800 & 151(4) & 7.4(5) & 28(9) \\
 & 5.30 & 0.136404 & 8600  & 178(4) & 9.0(6) & 12(2) \\
 & 5.32 & 0.136460 & 8040  & 185(4) & 8.9(5) & 10(2) \\
 & 5.33 & 0.136486 & 12000 & 187(4) & 7.2(6) & 14(3) \\
 & 5.34 & 0.136510 & 14280 & 190(4) & 8.1(4) & 14(2) \\
 & 5.35 & 0.136532 & 14320 & 193(5) & 8.9(5) & 16(3) \\
 & 5.36 & 0.136553 & 14200 & 196(5) & 9.4(4) &  9(1) \\
 & 5.37 & 0.136573 & 13440 & 199(5) & 8.7(5) &  9(2) \\
 & 5.38 & 0.136592 & 10520 & 203(5) & 9.0(5) &  9(2) \\
 & 5.39 & 0.136609 &  8200 & 206(5) & 8.2(6) &  9(2) \\
 & 5.40 & 0.136625 &  8440 & 209(5) & 8.5(5) &  8(1) \\
 & 5.45 & 0.136691 &  8560 & 227(5) & 5.9(7) &  8(2) \\
 & 5.50 & 0.136735 &  7840 & 245(6) & 5.6(6) &  5(1) \\
\hline
\end{tabular}
\end{center}
\caption{Run parameters of the simulations. We list the bare lattice coupling
$\beta$, the hopping parameter $\kappa$, the number of molecular dynamics units
(MDU), temperature, renormalised quark mass in the $\overline{\text{MS}}$-scheme
at a renormalisation scale of 2~GeV and the autocorrelation time of the
plaquette ($\tau_{U_p}$).}
\label{tab:scan-runparams}
\end{table}

\begin{table}[t]
\begin{center}
\footnotesize
\begin{tabular}{c|rr|rrr|rrr}
\hline
\hline
$T$ [MeV] & $\left< L \right>$ & $\left< L_S \right>$ & $\cond\bare$ &
$\bar{\chi}_\cond\bare$ & $\cond\ren$ & $\cond\bare_{\rm sub}$ &
$\bar{\chi}_{\cond_{\rm sub}}\bare$ & $\cond\ren_{\rm sub}$ \\
 & $\cdot 10^4$ & $\cdot 10^2$ & & $\cdot r_0^2$ & $\cdot r_0^3 \cdot 10^2$ &
& $\cdot r_0^2$ & $\cdot r_0^3 \cdot 10^2$ \\
\hline
\hline
\scan{B1}$_{\rm \kappa}$ & & & & & & & & \\
 201 & 4.6(6) & 3.1(3) & & & & & \\
 209 & 5.1(5) & 2.8(2) & & & & & \\
 218 & 6.8(4) & 3.7(3) & 0.269397(9) & 0.28(9) & -2.5(7) & 0.1691(13)
& 0.47(10) & -7.2(1.3) \\ 
 227 & 6.9(4) & 4.0(2) & 0.269085(6) & 0.21(5) & -1.8(5) & 0.1649( 6) &
0.52(11) & -8.0(9) \\
 236 & 10.0(4) & 5.6(2) & 0.268751(6) & 0.24(6) & -4.5(6) & 0.1592(5) &
0.27(11) & -12.9(8) \\
 238 & 9.3(6) & 5.0(3) & & & & & \\
 249 & 8.7(3) & 4.7(2) & 0.268614(7) & 0.21(5) & -3.6(7) & 0.1598(6) &
0.44(10) & -10.5(1.1) \\
 243 & 9.9(6) & 5.3(3) & 0.268534(5) & 0.22(5) & -4.3(5) & 0.1584(5) &
0.46(16) & -12.3(9) \\
 244 & 10.2(6) & 5.4(3) & & & & & \\
 245 & 9.9(7) & 5.3(3) & 0.268459(7) & 0.23(7) & -4.8(8) & 0.1586(7) &
0.42(6) & -11.3(1.3) \\
 246 & 11.3(5) & 6.1(2) & & & & & \\
 248 & 11.3(4) & 5.9(2) & 0.268378(6) & 0.13(4) & -5.9(7) & 0.1568(4) &
0.19(5) & -14.1(9) \\
 250 & 11.5(3) & 6.0(2) & 0.268304(8) & 0.18(4) & -6.5(9) & 0.1570(5) &
0.33(10) & -13.0(1.1) \\
 252 & 11.5(4) & 6.1(2) & & & & & \\
 255 & 12.9(3) & 6.5(2) & 0.268156(4) & 0.12(4) & -8.0(6) & 0.1551(3) &
0.15(3) & -16.1(1.0) \\
 265 & 13.5(5) & 6.9(2) & 0.267863(3) & 0.16(6) & -12.5(10) & 0.1527(1) &
0.04(2) & -20.7(1.4) \\
 275 & 15.9(4) & 7.7(2) & 0.267589(4) & 0.11(7) & -17.2(18) & 0.1515(1) &
0.08(3) & -23.9(2.1) \\
\hline
\scan{C1} & & & & & & & & \\
 151 & 0.7(4) & 0.59(3) & 0.272657(4) & 0.38(3) & -0.8(9) & 0.2849(8) & 2.23(20)
& -2.7(3) \\
 178 & 2.0(4) & 1.21(4) & 0.271101(3) & 0.34(3) & -2.3(2) & 0.2050(7) & 1.05(9)
& -6.3(4) \\
 195 & 4.9(3) & 2.76(6) & 0.270331(3) & 0.31(2) & -3.3(3) & 0.1831(5) & 0.59(7)
& -9.5(7) \\
 199 & 4.8(4) & 2.76(6) & 0.270120(3) & 0.24(2) & -4.1(3) & 0.1769(4) & 0.33(4)
& -10.5(8) \\
 203 & 5.7(4) & 3.52(6) & 0.269991(3) & 0.29(3) & -4.3(3) & 0.1723(4) & 0.26(4)
& -11.8(9) \\
 206 & 5.7(4) & 3.45(6) & 0.269869(2) & 0.21(2) & -4.1(3) & 0.1716(3) & 0.20(3)
& -11.6(9) \\
 209 & 5.4(4) & 3.40(6) & 0.269754(3) & 0.24(2) & -3.2(2) & 0.1724(4) & 0.33(4)
& -10.6(9) \\
 213 & 5.5(4) & 3.31(8) & 0.269627(3) & 0.29(2) & -3.9(3) & 0.1709(5) & 0.35(4)
& -11.1(10) \\
 216 & 6.9(4) & 3.85(6) & 0.269503(2) & 0.21(2) & -4.1(3) & 0.1679(3) & 0.15(2)
& -11.9(11) \\
 220 & 7.8(4) & 4.49(7) & 0.269375(2) & 0.16(2) & -4.9(3) & 0.1646(3) & 0.09(2)
& -12.8(11) \\
 223 & 8.1(4) & 4.80(7) & 0.269267(2) & 0.22(2) & -4.1(3) & 0.1657(4) & 0.20(3)
& -12.1(11) \\
 227 & 9.7(4) & 5.38(6) & 0.269133(2) & 0.13(2) & -5.7(3) & 0.1608(1) & 0.01(1)
& -13.7(10) \\
 245 & 11.6(4) & 6.21(7) & 0.268564(2) & 0.15(2) & -6.3(3) & 0.1574(3) & 0.09(4)
& -15.0(6) \\

\hline
\scan{D1} & & & & & & & & \\
 151 & 0.8(7) & 0.89(6) & 0.2725611(11) & 0.64(8) & -2.1(2) & 0.2859(50) &
3.89(76) & -6.1(4) \\
 178 & 2.5(4) & 1.89(6) & 0.2710744( 4) & 0.43(4) & -3.6(3) & 0.2068(17) &
1.54(32) & -9.7(6) \\
 185 & 3.3(3) & 2.15(6) & 0.2708011( 4) & 0.37(3) & -3.9(3) & 0.1963(11) &
0.67(12) & -10.6(7) \\
 187 & 3.1(3) & 2.47(6) & 0.2706568( 3) & 0.35(3) & -4.4(3) & 0.1909(10) &
0.71(15) & -11.1(9) \\
 190 & 3.4(3) & 2.36(5) & 0.2705272( 3) & 0.37(3) & -4.3(3) & 0.1901(10) &
0.93(29) & -10.8(9) \\
 193 & 4.0(3) & 2.78(5) & 0.2703920( 2) & 0.33(2) & -4.5(3) & 0.1850( 9) &
0.69(16) & -11.3(10) \\
 196 & 4.1(3) & 2.86(5) & 0.2702744( 3) & 0.32(3) & -3.9(3) & 0.1858(11) &
0.83(27) & -10.8(11) \\
 199 & 4.7(3) & 2.98(5) & 0.2701396( 2) & 0.30(2) & -4.3(3) & 0.1807( 7) &
0.42(7) & -11.3(12) \\
 203 & 4.9(3) & 3.14(6) & 0.2700152( 3) & 0.29(2) & -4.1(3) & 0.1778( 7) &
0.28(5) & -11.5(12) \\
 206 & 5.4(4) & 4.54(7) & 0.2698810( 3) & 0.24(2) & -4.7(3) & 0.1727( 7) & 
0.27(13) & -12.2(13) \\
 209 & 6.5(3) & 3.96(7) & 0.2697557( 3) & 0.23(2) & -4.8(3) & 0.1705( 6) & 
0.16(4) & -12.3(13) \\
 227 & 8.5(4) & 4.93(8) & 0.2691539( 2) & 0.15(2) & -5.2(3) & 0.1619( 3) &
0.02(7) & -13.2(12) \\
 245 & 10.8(4) & 5.85(9) & 0.2685848( 2) & 0.15(2) & -6.0(3) & 0.1580( 2) &
0.02(5) & -14.2(6) \\
\hline
\end{tabular}
\end{center}
\caption{Simulation results for the (smeared)  real part of the Polyakov
loop, the bare chiral condensate and its disconnected susceptibility
($\cond\bare$ and $\bar{\chi}_\cond\bare$), the associated renormalised
condensate ($\cond\ren$) in units of $r_0$, and the subtracted versions
($\cond\bare_{\rm sub}$, $\bar{\chi}_{\cond_{\rm sub}}\bare$ and $\cond\ren_{\rm
sub}$).}
\label{tab:scan-results}
\end{table}

\begin{table}[t]
\begin{center}
\small
\begin{tabular}{c|rrrr|rr}
\hline
\hline
$T$ [MeV] & $aM_P$ & $aM_S$  & $aM_V$ & $aM_A$ & $a\Delta M_{PS}$ & $a\Delta
M_{VA}\cdot 10^2$ \\
\hline
\hline
\scan{C1} & & & & & & \\
 151 & 0.136(2) & --- & 0.348(15) & 0.482(35) & --- & -12.42(12) \\
 178 & 0.127(2) & 0.201(38) & 0.311(13) & 0.370(33) & -0.102(25) & -4.94(56) \\
 195 & 0.153(4) & 0.164(31) & 0.332( 9) & 0.344(19) & -0.048(16) & -1.32(24) \\
 199 & 0.146(6) & 0.172(13) & 0.304(10) & 0.317( 9) & -0.041(11) & -0.70(21) \\
 203 & 0.186(5) & 0.225(17) & 0.351( 4) & 0.344( 8) & -0.049(13) & -0.47(13) \\
 206 & 0.189(4) & 0.210(15) & 0.324( 6) & 0.315(11) & -0.036( 9) & -0.57(12) \\
 209 & 0.164(5) & 0.220(30) & 0.332( 5) & 0.336(11) & -0.075(22) & -0.84(17) \\
 213 & 0.152(6) & 0.243(23) & 0.332( 6) & 0.331( 9) & -0.097(20) & -0.76(14) \\
 216 & 0.196(5) & 0.252(15) & 0.338( 5) & 0.341( 7) & -0.065(15) & -0.38(11) \\
 220 & 0.205(7) & 0.240(18) & 0.345( 4) & 0.343( 5) & -0.051(12) & -0.96( 9) \\
 223 & 0.180(7) & 0.234(41) & 0.336( 6) & 0.329(11) & -0.087(23) & -0.45(11) \\
 227 & 0.248(6) & 0.261( 7) & 0.359( 3) & 0.358( 4) & -0.014( 5) & { }0.01( 3)
\\
 245 & 0.264(6) & 0.282(10) & 0.356( 3) & 0.358( 3) & -0.028( 9) & -0.03( 3) \\

\hline
\scan{D1} & & & & & & \\
 151 & 0.104(3) & 0.365(107) & 0.349(21) & 0.412(25) & --- & -5.0(1.7) \\
 178 & 0.136(3) & 0.289(45) & 0.338(10) & 0.324(30) & -0.190(38) & -1.8(5) \\
 185 & 0.134(5) & 0.207(30) & 0.322( 9) & 0.323(25) & -0.086(28) & -1.8(4) \\
 187 & 0.150(5) & 0.142(29) & 0.339(14) & 0.375(12) & -0.037(13) & -0.5(3) \\
 190 & 0.136(4) & 0.150(19) & 0.310( 9) & 0.333(15) & -0.041(12) & -1.0(3) \\
 193 & 0.142(4) & 0.145(28) & 0.319( 8) & 0.342(12) & -0.031(15) & -0.4(2) \\
 196 & 0.136(4) & 0.186(37) & 0.311( 7) & 0.315(12) & -0.080(28) & -0.9(2) \\
 199 & 0.155(5) & 0.271(44) & 0.321( 7) & 0.323(14) & --- & -0.9(3) \\
 203 & 0.161(5) & 0.222(27) & 0.332( 6) & 0.324( 9) & -0.068(24) & -0.5(2) \\
 206 & 0.173(9) & 0.272(30) & 0.324( 4) & 0.326( 8) & -0.104(22) & 0.0(2) \\
 209 & 0.170(6) & 0.218(21) & 0.328( 6) & 0.349( 7) & -0.061(19) & -0.3(2) \\
 227 & 0.239(6) & 0.251(15) & 0.343( 4) & 0.349( 5) & -0.039(21) & -0.1(2) \\
 245 & 0.220(7) & 0.223(10) & 0.347( 4) & 0.348( 4) & -0.012( 6) & 0.1(1) \\
\hline
\end{tabular}
\end{center}
\caption{Simulation results for screening masses $M$ in $P$, $S$, $V$ and $A$
channels and the direct measurements for screening mass differences $\Delta M$.
The results for scan \scan{B1}$_{\rm \kappa}$ are not listed.}
\label{tab:scm-results}
\end{table}

\clearpage

\bibliographystyle{JHEP}
\bibliography{spires}

\end{document}